\documentclass[a4paper,11pt]{article}

\usepackage[utf8]{inputenc}
\usepackage[right=2cm,left=2cm,top=2cm,bottom=3cm]{geometry}
\usepackage{amsmath,amsfonts,amssymb,bm}
\usepackage{mathtools}
\usepackage{booktabs}
\usepackage[colorlinks,linkcolor=blue,citecolor=blue,urlcolor=blue]{hyperref}
\usepackage{dsfont}	
\usepackage{listings}
\usepackage{xparse}

\NewDocumentCommand{\codeword}{v}{%
\texttt{\textcolor{black}{#1}}%
}

\usepackage{natbib}
\usepackage[colorlinks,linkcolor=blue,citecolor=blue,urlcolor=blue]{hyperref}
\usepackage{tikz}       
\usetikzlibrary{bayesnet}     
  
\usepackage{forest}
\usetikzlibrary{arrows.meta, shapes.geometric, calc, shadows}

\colorlet{grey}{black!20}
\colorlet{black}{black}
\colorlet{white}{white}

\makeatletter
\newcommand\footnoteref[1]{\protected@xdef\@thefnmark{\ref{#1}}\@footnotemark}
\makeatother
\usepackage{DejaVuSansMono}
\usepackage[scaled=0.8]{beramono}

 \usepackage{array}
 \newcolumntype{L}[1]{>{\raggedright\let\newline\\\arraybackslash\hspace{0pt}}m{#1}}
\newcolumntype{C}[1]{>{\centering\let\newline\\\arraybackslash\hspace{0pt}}m{#1}}
 \newcolumntype{R}[1]{>{\raggedleft\let\newline\\\arraybackslash\hspace{0pt}}m{#1}}

\newcommand{\MU}{\bm{\mu}}

\newcommand{\Y}{\mathbf{Y}}

\newcommand{\z}{\mathbf{z}}

\newcommand{\arch}{y_{ij}^{(k)} }
\newcommand{\ak}{\alpha^{(k)} }
\newcommand{\bk}{\beta^{(k)} }

\usepackage{authblk}
\title{\textbf{Model-based clustering for multidimensional social networks}}
\date{}
\author[1]{Silvia D'Angelo}
\author[2]{Marco Alfò}
\author[1]{Michael Fop}
\affil[1]{School of Mathematics and Statistics, University College Dublin, Ireland}
\affil[2]{Department of Statistical Sciences, La Sapienza University of Rome, Italy}
\date{}                     
\begin{document}
\maketitle



\begin{abstract}
 Social network data are relational data recorded among a group of actors, interacting in different contexts. 
 Often, the same set of actors can be characterized by multiple social relations, captured by a multidimensional network. A common situation is that of colleagues working in the same institution, whose social interactions can be defined on professional and personal levels. 
 In addition, individuals in a network tend to interact more frequently with similar others, naturally creating communities.
 Latent space models for network data are useful to recover clustering of the actors, as they allow to represent similarities between them by their positions and relative distances in an interpretable low dimensional social space.
 We propose the infinite latent position cluster model for multidimensional network data, which enables model-based clustering of actors interacting across multiple social dimensions. The model is based on a Bayesian nonparametric framework, that allows to perform automatic inference on the clustering allocations, the number of clusters, and the latent social space.
  The method is tested on simulated data experiments, and it is employed to investigate the presence of communities in two multidimensional networks recording relationships of different types among colleagues.  
\end{abstract}
 

\section{Introduction}
\label{intro}
Network data describe relations and interconnections among $n$ units. Interacting units are denoted as \emph{nodes}, while connections are called \emph{edges}. Well known examples are social network data, where friendship, approval, admiration, and other social relations are expressed between individuals and may be modelled to understand how people interact in the society. In general, connections observed in network data are of difficult visualization and interpretability, due to the complex nature of networks themselves. Therefore, network analysis methods mainly focus on reconstructing and explaining the connections observed among the nodes. Such methods have to deal with the double task of faithfully modelling the relations between the units, which lead to the observed connections, and of providing with a feasible and interpretable summary of the data. Observed connections are modelled stochastically and each pair of nodes is associated to an edge probability. Edge probabilities describe dyads connection propensity and can be modelled in many different ways, according to the data at hand or the purpose of the analysis.

First attempts to model network data were built on the assumption of independence between the edges. Edge probabilities were assumed to be constant, see \citep{rg}, or simple functions of some network statistics as in \citep{p1,p2}. These first models paved the way for a more in depth analysis of edge formation in network data, where the edges independence assumption was either reduced or removed and complexity was added to the specification of edge probabilities \citep{wasserPatti,robin2006,sto_bloc,snj,nowicki:2001,hoff}. Among these models, the class of latent variable models has gained a growing attention in the last years. 
Such models are particularly interesting and meaningful as they explain the observed interconnection structure in network data by means of latent variables, which capture the association between the nodes.
First latent variable models are the latent space model by \cite{hoff} and the stochastic block model by \cite{sto_bloc,snj}. The latter may be thought as a latent class analysis model for network data and it is explicitly designed to model clustering of the nodes. Clustering is a feature often observed in many real world social network data, as actors may tend to interact more frequently within the community they belong to. 
The stochastic block model classifies the nodes into different clusters and provides a framework for modelling between cluster interaction.
Some recent developments on the topic are those by \cite{signorelli:2018}, which analyses bill co-sponsorships in the Italian Chamber of Deputies, \cite{Bouv:2018}, which extends the stochastic block model to the analysis of textual data, and those by \cite{Matias:2018,bartolucci:2018, rastelli:2020}, extending the model to time-dependent network data. Although stochastic block models may synthetically represent clusters, they fail to represent within cluster transitivity, which refers to how nodes locally interact.
A flexible extension of the stochastic block model which addresses such issue is the mixed membership stochastic block model by \cite{mm_sto_bloc}. In this framework, nodes may belong to different clusters, depending on whom they are interacting with. 
A different approach is that of model-based clustering for latent space models, introduced by \cite{mbcsn}. This framework is based on the class of latent space models, and directly accounts for transitivity in social network data \citep{hoff,mbcsn}. Edge probabilities are described as a function of node positions in an unobserved space, which is responsible for the observed structure in the social network. Such latent positions arise from a mixture distribution, whose components correspond to clusters in the data, as in standard model-based clustering \citep{Fraley:2002}. 
The original model is estimated with a MCMC algorithm; \cite{Salter:2013} re-implemented it with a variational Bayesian inference approach. An extension to the framework of latent position cluster model is that by \cite{Gormley:2010}, which combines it with a mixture of experts framework.
\cite{Fosdick:2018} attempt to bridge stochastic block models and latent position cluster models, using the so-called \emph{Latent Space Stochastic Blockmodel}. In this framework, within cluster probabilities are modelled via a latent space model, while between cluster interactions are expressed as in stochastic block models.

In many situations, the same set of nodes can be characterized by multiple social relations. For example, social interactions among colleagues working in the same institution often happen on a professional level (like working on the same project or asking for advice on a given matter) and on a personal level (like being friends or participating to the same recreating activity). Recording these multiple interactions between the same actors generates a complex type of network data, referred to as \emph{multidimensional networks}, or \emph{multiplexes}. A multiplex is a collection of networks, often referred as \emph{views}, where multiple sets of edges are observed for the same set of nodes. Latent space models for multiplex data have been recently proposed by \cite{gollini, Salter:2013, mio}, which  parsimoniously model the collection of social interactions using either a single or multiple latent space representations. Despite multiplexes having often actors being grouped into communities, to the best of our knowledge, only few works have been proposed for clustering the nodes of multidimensional networks. Methods based on stochastic block models are those proposed by \cite{Ishiguro:2010,barbillon:2017, Subhadeep:2020}.
Instead, within the class of latent space methods \cite{sewell3} extended the work of \cite{mbcsn} for clustering longitudinal network data, a particular case of multiplex data, where the multiple views refer to a single social relation recorded at different time occasions. This work relies on a fixing the number of clusters, selected using a model selection criterion.

In general, in model-based clustering, The number of clusters is often unknown and needs to be inferred from the data. 
The problem of selecting the number of clusters is often reduced to selecting the number of mixture components. Often, this problem is recast as a model selection problem, whereby different model selection criteria and procedures have been proposed to select the number of clusters, for a review see \cite{Mclachlan:2014}. These approaches often require estimating and comparing models for multiple choices of the number of components, which can be computationally demanding and even unfeasible in some applications. 
In model-based clustering within the Bayesian framework, reversible jump MCMC algorithms \citep{Green:1995} allow to automate the inference on the unknown number of components, as posterior samples are drawn from a space of varying dimensions \citep{Zhang2004}. However, trans-dimensionality makes reversible jump algorithms quite computationally intensive, which often requires ad-hoc implementations. 
Another approach for inference on the number of clusters is that of overfitting mixture models by \cite{malsiner:2016}. 
Sparse hierarchical priors are placed on the mixture weights and component parameters, resulting in a mixture model where superfluous components are emptied during MCMC sampling. This approach estimates the number of components as the most frequent number of non-empty components visited during MCMC sampling. However, it requires the specification of an upper bound on the number of components. Related to the overfitting mixture distribution approach is that of Dirichlet process mixture models \citep{Escobar:1995,Rasmussen:2000,frufru:2018}.
Here, the number of components is assumed to be potentially infinite, and the distributions of component-specific parameters are taken to be the realizations of a Dirichlet process \citep{Antoniak:1974}. Since in the observed sample the number of components is finite, joint inference on the clustering allocations, the number of components, and the mixture distribution parameters is made possible by the Dirichlet process specification. Differently from overfitting mixture models, no upper bound on the number of components is required, and the number of clusters is allowed to change with the sample size.

In this work, we introduce the \emph{infinite latent position cluster model} for model-based clustering of the nodes of multidimensional networks.
The proposed framework is built within the class of latent space models and allows to jointly estimate clustering allocations, cluster parameters, and latent coordinates without previous specification of the number of clusters. The contribution of our paper is two-fold: generalization of the model by \cite{mbcsn} to multidimensional networks, building on the work by \cite{mio}, and development of a Bayesian nonparametric model formulation, which allows automatic inference on the number of communities, made dependent on the number of actors in the social multiplex.

The paper is structured as follows: Section \ref{modello} presents the infinite latent position cluster model, after a brief overview of the main aspects of latent position models for networks. The model is developed within a hierarchical Bayesian framework, and Section \ref{estimation} illustrates the principal aspects of the MCMC algorithm implemented for inference. Some practical implementation details are discussed in Section \ref{further}. A simulation study to investigate the ability of the proposed model to recover the latent space and the clustering structure of the nodes is presented in Section \ref{simulation}. Section \ref{data} presents two illustrative data examples, both concerning the analysis of professional and personal relations between employers working in two different institutions. The first concerns the analysis of a multiplex describing the interactions among attorneys working in the same law firm; the second considers clustering of the members of a Computer Science department. 
The paper ends with a discussion in Section \ref{discussion}.

\section{A latent position cluster model for multidimensional networks}
\label{modello} 
In this section we introduce the {\em infinite latent position cluster model} (ILPCM), a generalization of the latent position cluster model by \cite{mbcsn}. Before introducing the ILPCM, we briefly review the main aspects of latent position models, pointing out some critical features.

\subsection{Latent position cluster model}
\label{modello:hoff} 
Let $\mathbf{Y}$ be the binary adjacency matrix of dimension $n \times n$, where $n$ is the number of nodes in the network. The entries of $\Y$ are such that $y_{ij} = 1$ if nodes $i$ and $j$ are connected, 0 otherwise. In the context of unidimensional network data, the {\em latent position model} (LPM) of \cite{hoff} assumes that each node has a corresponding position in a $p$-dimensional Euclidean social latent space. The probability of a tie between any two nodes $i$ and $j$ is then expressed as a logistic function dependent of the distance between the corresponding latent coordinates $\z_i$ and $\z_j$, and additional parameters: 
\begin{equation}
    \Pr\Bigl(y_{ij}= 1 \mid \alpha, \beta , \z_i, \z_j \Bigr) =\frac{\exp\bigl\{ \alpha -\beta  \lVert\z_i - \z_j \rVert_2^2 \bigr\} }{1 + \exp\bigl\{ \alpha -\beta \lVert\z_i - \z_j \rVert_2^2 \bigr\} },
    \label{eq:singleN}
\end{equation}
where $\alpha$ is an intercept capturing the overall connectivity in the network, and $\beta$ $(\beta \geq 0)$ is a scale parameter which weights the influence of the latent space on the edge probabilities; when $\beta\approx 0$, the latent space is practically irrelevant and the edge probabilities reduce to those generated by a random graph \citep{rg,mio}. The distance relates the latent coordinates to the probability of a tie, so that nodes closer in the latent space are more likely to be connected in the observed social network. In this model formulation we take the distance to be the squared Euclidean distance \citep[see][]{gollini,mio}, although different distances and geometries of the latent space could be considered \citep{hoff,smith:2019}.

To account for a clustering structure of the nodes, \cite{mbcsn} introduced the {\em latent position cluster model} (LPCM). This model enables model-based clustering for social networks, extending the LPM by assuming that the latent coordinates arise from a finite mixture of $G$ spherical $p$-variate Gaussian distributions:
\[
\mathbf{z}_i \thicksim \sum_{g=1}^G \pi_g \, \mathcal{N}_p \bigl(\bm{\mu}_g, \sigma_g^2 \mathbf{I}\bigr),
\]
where $\pi_g$ are the mixture weights, such that $\pi_g > 0 $, $g=1,\dots,G$, and $\sum_{g=1}^G \pi_g  = 1$; $\bm{\mu}_g$ and $\sigma_g^{2}$ are the component-specific means and variances. 

A distinctive aspect of the LPCM is the parametric nature of the mixture distribution, where the specification of a finite number of components $G$ generates a difficult model selection problem, corresponding to the selection of the number of clusters. 
To the best of the authors knowledge, the problem of inferring the number of communities when clustering network data using latent space type models has only been scarcely explored. In \cite{mbcsn} the authors address this issue by using a version of the Bayesian Information Criterion \citep[BIC,][]{BIC}, commonly used as model selection tool in model-based clustering \citep{Mclachlan:2014,Fraley:2002}.
To select $G$, the authors propose a pragmatic approach in which the marginal likelihood is approximated by conditioning on the estimated mode of the latent coordinates. \cite{Ryan:2017} point out that this method does not allow exploration of the posterior uncertainty on the number of clusters and several factors could influence the quality of the approximation and ultimately the results. In addition, the BIC approach poses a serious computational burden in many applications, since multiple models needs to be estimated, one for each pre-specified possible value of $G$. The authors propose an alternative Bayesian model selection method for the LPCM based on integrating out the model parameters, which allows efficient posterior inference on the number of components. Other works whose focus is the development of alternative latent space models based on finite mixture distributions employ various versions of standard model selection criteria \citep[see for example][]{Gormley:2010, sewell3, Ng:2021}. 

Bayesian nonparametric mixture models offer a flexible framework for inference on the clustering, assuming that the number of mixture components is infinite. Differently from a standard finite mixture model approach, like the LPCM, this avoids the prior specification of the number of components, or the specification of a range of number of clusters to perform model selection on. Instead, a nonparamteric approach permits automatic inference on the number of clusters, which is also allowed to grow as the number of actors in the network  grows \citep{muller:2015}.

To overcome the issue of number of cluster selection, we introduce a latent space approach for multidimensional network data based on a Bayesian nonparametric framework. Through a nonparametric prior, we incorporate selection of the number of clusters directly into our inference procedure, therefore avoiding the usually computationally intensive model selection step performed in LPCM and related models. We propose the use of a Dirichlet process mixture that leads to the formulation of an infinite mixture model for the latent coordinates \citep{Escobar:1995,Rasmussen:2000,muller2013,muller:2015}. The idea was only scantly explored for unidimensional networks in the discussion to the \cite{mbcsn} paper, here we propose a generalization to multidimensional networks. 


\subsection{The infinite latent position cluster model}
\label{modello:mio} 

Let $\mathbf{Y}=\{\mathbf{Y}^{(1)}, \ldots, \Y^{(k)}, \ldots \mathbf{Y}^{(K)} \}$ be a multiplex composed of $K$ networks (views). Each element $\mathbf{Y}^{(k)}$ is a binary adjacency matrix of dimension $n \times n$, where $n$ is the number of nodes in the multiplex. Recall that in multidimensional network data the number of nodes is constant across the views, while the observed edges may change with the view. The general element of $\mathbf{Y}^{(k)}$ is such that $y_{ij}^{(k)}= 1$ if nodes $i$ and $j$ are connected in the $k^{\text{th}}$ view, while $y_{ij}^{(k)}=0$ if they are not. Clearly, when $K=1$, the multidimensional network reduces to a single network with $n$ nodes. 

Following \cite{mio}, as in the LPM and LPCM, we assume that each node has an unknown position in a latent $p-$dimensional Euclidean space, whereby the probability of observing a connection between nodes $i$ and $j$ in one of the views is a function of their latent coordinates $\z_i$ and $\z_j$ and view-specific parameters:
\begin{equation}
 \Pr\Bigl( \arch = 1 \mid \alpha^{(k)}, \beta^{(k)} , \z_i, \z_j \Bigr) =\frac{\exp\bigl\{ \ak -\bk  \lVert\z_i - \z_j \rVert_2^2 \bigr\} }{1 + \exp\bigl\{ \ak -\bk \lVert\z_i - \z_j \rVert_2^2 \bigr\} }.
 \label{eq:prob1}
\end{equation}
The view-specific parameters $\ak$ and $\bk$ modify the effect of the latent space on network-specific edge probabilities and have the same interpretation of those in the LPM and LPCM for a single network. As in \cite{mio}, the distance scale coefficients $\bk$ are lower-bounded by  $0$, while the network intercepts $\ak$ are lower-bounded by $\log \Bigl( \frac{\log(n)}{n -\log(n)}\Bigr)$. The distance between any two latent coordinates is the squared Euclidean distance. Although we develop our proposal considering the squared Euclidean distance, the proposed framework may be extended to incorporate different specifications for the distance function as well.

Differently from other modeling approaches for multiplex data \citep[][for example]{gollini,salter}, we assume the existence of a unique latent space common to all the network views. 
This is intended to capture the complex dependences between the nodes in a parsimonious way \citep{mio}, embedding at the same time the potential presence of communities. To take into account the clustering structure, we assume that the node-specific latent coordinates are distributed according to a mixture of $p$-variate Gaussian distributions, but according to a data generating process that is different from that one of LPCM. We assume a Dirichlet process mixture distribution on the component parameters \citep{Ferguson:1973,Antoniak:1974,BayesianNonparametrics:2010,muller:2015}:
\begin{equation}
  \mathbf{z}_i \mid \bm{\mu}_i , \bm{\Sigma}_i \thicksim \mathcal{N}_p \bigl( \bm{\mu}_i, \bm{\Sigma}_i \bigr), \qquad \bm{\Omega}_i \thicksim S, \qquad S \thicksim DP\bigl( \psi,  S_0);
  \label{eq:dirEsteso}
\end{equation}
where $S_0$ is a continuous base distribution corresponding to the expected value of the process, and $\psi\geq 0$ is a concentration parameter. The set $\bm{\Omega}_i = ( \bm{\mu}_i, \bm{\Sigma}_i )$ includes the mean and covariance parameters of the distribution of the latent coordinates, distributed according to $S$ for $i=1, \dots, n$. Even though $S_0$ is continuous, distributions drawn around it are almost surely discrete. The degree of discretization depends on the concentration parameter $\psi$: the lower the value of $\psi$, the lower the number of unique realizations. The discrete nature of the Dirichlet process is what makes it suitable to describe mixture models. Indeed, in a finite sample of $n$ nodes, the number of unique realizations is finite and it can be denoted by $G \leq n$. This implies that some units come from a common component, that is some of the latent coordinates $z_i$ share the same prior distribution parameters $\Omega_g$, $g=1, \dots, G$. The set of parameters $\bm{\Omega}_g = (\bm{\mu}_g, \bm{\Sigma}_g)$, $g=1, \dots, G$, corresponds to the set of component parameters, and the number of realizations $G$ is indeed the number of mixture components. 

The Dirichlet process mixture on the component parameters induces the following infinite mixture distribution on the latent coordinates \citep{Rasmussen:2000}:
\begin{equation}
   \mathbf{z}_i \thicksim \sum_{g = 1}^{\infty} \pi_g \, \mathcal{N}_p \bigl( \bm{\mu}_g, \bm{\Sigma}_g \bigr). 
\label{eq:prioriZ}
\end{equation}
This model specification allows for joint estimation of the unknown number of components $G$ and the component-specific parameters $\bm{\mu}_g$ and $\bm{\Sigma}_g $. Such an infinite mixture distribution framework is particularly suited for network data, as the number of mixture components is bounded by the number of nodes \citep{Antoniak:1974}. Hence, it is implicitly assumed that the number of mixture components can vary if new nodes join the multiplex. This assumption is reasonable with network data, as new actors may alter the community structure depending on how they interact with the pre-existing nodes.

\section{Model estimation}
\label{estimation}
Here we provide the main aspects of the estimation procedure for the most general case of multiplex data ($K > 1$), although we note that the same considerations apply in the case of a single network ($K=1$).

As in standard latent position models, the presence/absence of a link between node $i$ and $j$ is assumed to be independent of all other connections in the network, conditionally on the latent coordinates of the two nodes. Based on the edge probability in equation \ref{eq:prob1}, we write the log-likelihood as follows:
\begin{equation}
    \ell \bigl(\mathbf{Y} \mid \bm{\alpha}, \bm{\beta}, \mathbf{Z}\bigr) = \sum_{k=1}^K \sum_{i = 1}^n \sum_{j\neq i} \arch(\ak -\bk \lVert\z_i - \z_j \rVert_2^2) -\log(\ak -\bk \lVert\z_i - \z_j \rVert_2^2),
\label{eq:loglik}
\end{equation}
where $\bm{\alpha}=(\alpha^{(1)}, \dots , \alpha^{(K)})$, $\bm{\beta}=(\beta^{(1)}, \dots , \beta^{(K)})$, and $\mathbf{Z}$ is the matrix of the latent coordinates. 

We propose a hierarchical Bayesian framework for inference. We assume the following prior and hyper-prior distributions on the logit parameters:
\[
\ak \thicksim \mathcal{N}_{[LB(\alpha), \infty]} \bigl( \mu_{\alpha} , \sigma_{\alpha}^2 \bigr), \quad \text{where:} \quad \mu_{\alpha}\mid \sigma_{\alpha}^2 \thicksim \mathcal{N}_{[LB(\alpha), \infty]} \bigl( m_{\alpha} , \tau_{\alpha}\sigma_{\alpha}^2 \bigr), \quad \sigma_{\alpha}^2\thicksim \text{Inv} \chi_{\nu_{\alpha}}^2;
\]
and
\[
\bk \thicksim \mathcal{N}_{[0, \infty]} \bigl( \mu_{\beta} , \sigma_{\beta}^2 \bigr), \quad \text{where:} \quad \mu_{\beta}\mid \sigma_{\beta}^2 \thicksim \mathcal{N}_{[0, \infty]} \bigl( m_{\beta} , \tau_{\beta}\sigma_{\beta}^2 \bigr), \quad \sigma_{\beta}^2\thicksim \text{Inv} \chi_{\nu_{\beta}}^2;
\]
for $k=1, \dots,K$, where $\mathcal{N}_{[a, \infty]}$ denotes a Normal distribution truncated at $a$. Recall that the lower bound for $\ak$ is $LB(\alpha) = \log \Bigl( \frac{\log(n)}{n -\log(n)}\Bigr)$. Since the estimation of $\bm{\alpha}$ and $\bm{\beta}$ does not directly depend on the latent coordinates, we proceed using the same proposal and full conditional distributions derived in \cite{mio}; see the Appendix section of \cite{mio} for further details.

Differently from the LPCM in \cite{mbcsn} (see Section \ref{modello:hoff}), we opt for a more flexible specification of the component-specific covariance matrices. We let the component variances vary across the latent dimensions, assuming non-spherical but diagonal covariance matrices: $\bm{\Sigma}_g = \mathrm{diag}(\sigma_{1g}^2, \ldots, \sigma_{rg}^2 \ldots \sigma_{pg}^2)$. This specification allows to capture heterogeneity of the nodes across the latent dimensions and the components. 
The idea is that each latent coordinate add a further information which is not yet gathered by the other ones. Further, the volume of information is coordinate-specific, as some coordinates may be more spread than others.
Given the diagonal form of the component covariance matrices, we specify $S_0$ as a Normal-Inverse Gamma distribution:
\[
\bigl( \mu_{rg}, \sigma_{rg}^2 \bigr)\thicksim NIG \bigl( m_r, \tau_z \sigma_g^2 , \nu_1, \nu_2\bigr) , \quad r = 1, \dots,p,
\]
where $m_r, \tau_z, \nu_1,$ and  $\nu_2$  (for all $g=1, \dots,G$) represent the mixture components hyperparameters.

The estimation of the latent coordinates and related mixture component parameters is not that straightforward. The Dirichlet process mixture formulation of the ILPCM in equations \eqref{eq:dirEsteso} and \eqref{eq:prioriZ} requires special care.
Given a sample of dimension $n$ and a number of groups $G \leq n$, the allocation of the network nodes to the clusters is unknown.
We introduce an auxiliary multinomial cluster membership variable $\bm{c} = (\bm{c}_1, \ldots, \bm{c}_{i},\ldots,  \bm{c}_n)$. The $i^{\text{th}}$ entry is a $G$-dimensional binary vector $ \bm{c}_i= (c_{i1}, \ldots, c_{ig}, \ldots, c_{iG})$, such that $\sum_{g=1}^G c_{ig} = 1$, and whose $g^{\text{th}}$ entry is 1, indicating if the $i^{\text{th}}$ latent coordinate comes from the $g^{\text{th}}$ component. Using the membership indicator $\bm{c}_i$, we can write the distribution of latent coordinates and cluster allocations for each node as follows:
\begin{equation}
    \bigl( \mathbf{z}_i, \bm{c}_i\bigr) \mid {\bm{\mu}_g, \bm{\Sigma}_g } \thicksim \prod_{g=1}^G \Bigl[ \pi_g \mathcal{N}_p\, \bigl(\bm{\mu}_g, \bm{\Sigma}_g \bigr)\Bigr]^{c_{ig}}.
\label{eq:completeZ}
\end{equation}
Hence, given the mixture representation above, the mixing weights can be simply rewritten as $\pi_g = n^{-1}\sum_{i = 1}^n c_{ig}$.
The unknown quantities of interest are now the actual number of components $G$, the component-specific parameters $\bm{\Omega}_g =(\bm{\mu}_g, \bm{\Sigma}_g)$, $g=1, \dots,G$, and the cluster allocations $\bm{c}_i$, $i = 1, \dots,n$. To estimate such quantities, we exploit the \emph{Chinese restaurant} representation of the Dirichlet process \citep{Aldous:1985}. Inference procedures for the Chinese restaurant representation of the Dirichlet process when the base distribution is a Normal-Inverse Gamma have been widely studied in the literature, see \cite{BayesianNonparametrics:2010,muller:2015}. In particular, we adopt the proposal by \cite{Bush:1996} to update the cluster labels $\bm{c}_i$. Furthermore, we assume that the concentration parameter of the Dirichlet process has a Gamma prior distribution, $\psi \thicksim \Gamma\bigl(\xi_1, \xi_2 )$, and we adopt the proposal by \cite{Escobar:1995} to update this parameter. 


Given an observed multidimensional network $\mathbf{Y}$ with $n$ nodes and $K$ views, we propose a Metropolis within Gibbs MCMC algorithm for inference. The algorithm sequentially updates values for the logit parameters $\bm{\alpha}$, and $\bm{\beta}$, the corresponding nuisance parameters, the latent coordinates, the component parameters, and the group labels. The update steps of the logit parameters and related nuisance parameters follow those described in \cite{mio}, and we refer the reader to this article for details. The update steps of the latent coordinates, mixture parameters, and cluster allocations are briefly described below; further details can be found in Appendix~\ref{appendice3}:
\begin{enumerate}
     \item  The latent coordinates $\mathbf{z}_i$ are updated sequentially from the proposal distribution presented in Appendix \ref{appendix:latent_coordinates} for each node in turn. 
    To account for rotation and translation invariance of the latent space, we compare the set of latent coordinates obtained at the previous iteration with the updated one, computing their Procrustes correlation. If the value of this correlation is greater than a pre-specified threshold, the updated set of coordinates is discarded and the one at the previous iteration is kept.
    \item The cluster allocations $\bm{c}_i$, $i=1, \dots,n$, are updated from their full conditional distribution, whose specification is given in Appendix \ref{appendix:labels}. Mixture weights $\pi_g$ are simply computed as $n_g/n$, where $n_g$ is the number of nodes assigned to cluster $g$.
    \item The mixture component parameters $\bm{\mu}_g, \bm{\Sigma}_g$ are updated using the full conditional distributions provided in Appendix~\ref{appendix:param}. 
    \item The Dirichlet concentration parameter $\psi$ is updated using the proposal described in Appendix~\ref{appendix:concentration}
\end{enumerate}

Section \ref{simulation} presents a simulation study to evaluate the performance of the proposed estimation procedure.

\section{Practical implementation details}
\label{further}
The ILPCM and the MCMC estimation procedure proposed in Section \ref{estimation} present some practical aspects that we address in detail in the following sections. 

\subsection{Logit parameters identifiability}
\label{identif}
To guarantee identifiability, a network in a given multiplex has to be taken as a reference network, and the corresponding parameters $(\ak,\bk) = (\alpha^{(\text{ref})}, \beta^{(\text{ref})})$ need to be fixed. We suggest to set $\beta^{(\text{ref})} =1$. This constraint does not alter the interpretation of the scale coefficient parameters $\bk$, as their values are meaningful only when compared with each other. We propose to choose $\alpha^{(\text{ref})}$ as suggested in \cite{mio}:
\begin{equation}
  \alpha^{(\text{ref})} \geq \log \left( \frac{\bar{p}}{1-\bar{p}}\right) +2,
\label{eq:alpharef}  
\end{equation}
where $\bar{p} = \sum_{i=1}^n \sum_{j \neq i} y_{ij}^{(\text{ref})}/ \bigl[ n(n-1) \bigr]$ denotes the empirical density of the reference view of the multiplex. The term $2$ on the right side of equation \ref{eq:alpharef} is the mean empirical distance among coordinates simulated from a standard Gaussian distribution. In the present model, we assume that the prior distribution for the latent coordinates is a mixture of Gaussian distributions with unknown number of components. Therefore, it is not possible to empirically estimate the average distance among coordinates. However, it is reasonable to expect that coordinates drawn from a mixture of Gaussian distributions will be on average further apart than coordinates drawn from a single Gaussian distribution. The $\geq$ condition in equation \ref{eq:alpharef} derives from this last consideration.

\subsection{Initialization}
\label{init}
We adopt different procedures to initialize the model parameters and the latent space in the proposed MCMC algorithm (Section \ref{estimation}). For an observed multiplex, geodesic distance matrices are computed from the networks. Then, starting values for the latent coordinates $\bm{z}$ are obtained via multidimensional scaling on the average geodesic distance matrix, given a chosen value of the latent space dimension $p$ \citep{hoff, mio}. The initial partition of the nodes is obtained by fitting a ``VVI'' mixture model on the starting $\bm{z}$ coordinates, using the \codeword{mclust} R package \citep{Scrucca:2016} and retaining the corresponding optimal classification selected by BIC. Subsequently, distances between the nodes are initialized by computing the squared Euclidean distance between the starting $\bm{z}$ coordinates. These distances are then used to fit $K$ logistic regressions, where each view adjacency matrix is vectorized and used as response variable. The corresponding intercept and scaling coefficient estimates are set as initial values of $\ak$ and $\bk$, $k=1,\dots,K$, respectively. Last, $(\mu_{\alpha}, \mu_{\beta}, \sigma_{\alpha}^2, \sigma_{\beta}^2)$ are initialized as the mean and variance of the starting values of $\ak$ and $\bk$, respectively.

\subsection{Posterior distributions post-processing }
\label{postproc}

Throughout the iterations, due to the Bayesian nonparametric formulation of the ILPCM, the MCMC algorithm explores mixture models with different number of components. This leads to the problem of how to summarize the posterior of the cluster allocations and mixture component parameters, and derive a unique point estimate of the number of clusters, of the clustering allocation of the network nodes, and of the component parameters. 

In Bayesian nonparametrics, and in general Bayesian estimation of mixture models, posterior inference on the cluster assignments is a difficult task. Multiple approaches have been proposed in the literature, see \cite{fruhwirth:2019} for an overview. 
Here, to estimate the cluster allocation of the nodes and consequently the number of clusters given the data, we adopt the procedure of \cite{wade:2018}. This procedure is based on the estimated posterior similarity matrix, containing the proportion of times a pair of nodes $(i,j)$ are in the same cluster. The method provides the optimal partition, obtained by minimization of the Variation of Information loss, computed using the estimated posterior similarity matrix. The estimated $\hat{G}$ value is retrieved as the number of clusters composing the optimal partition. 

A similar complication arises when summarizing the posterior distributions of the component parameters $\bm{\Omega}_g =(\bm{\mu}_g, \bm{\Sigma}_g)$, $g=1,\dots,G$. 
Different iterations of the algorithm present a different number of component-specific parameters $\bm{\Omega}^{(t)} = \bigl(\bm{\Omega}_1^{(t)},\dots, \bm{\Omega}_g^{(t)}, \dots, \bm{\Omega}_{G^{(t)}}^{(t)}\bigr)$, where $G^{(t)}$ is the number of mixture components at iteration $t$. However, after having obtained an estimate $\hat{G}$ of the number of components, precisely $\hat{G}$ parameter estimates are needed to describe the mixture distribution. Taking in consideration only the posterior samples of the component-specific parameters for which $G^{(t)} = \hat{G}$ would not take into account the information and uncertainty carried by those samples whose number of components is different from the a posteriori estimate $\hat{G}$. For this reason, we adopt the procedure suggested by \cite{frufru:2011}. The author proposes a k-means based approach to summarize the whole posterior distributions of the mixture component parameters $\MU_g, \Sigma_g$ for a given estimate of the number of clusters. K-means clustering with $\hat{G}$ clusters is applied on the whole collection of posterior samples (after burn-in) of the component means, obtaining a classification of the posterior draws into $\hat{G}$ clusters. A permutation test is employed to order the draws and ensure a unique labeling. The same k-means classification is then used to reorder the posterior draws of the component covariance matrices. Finally, estimates of the cluster parameters $\hat{\bm{\Omega}}_g = (\hat{\bm{\mu}}_g, \hat{\bm{\Sigma}}_g)$ are obtained as the Monte Carlo averages of the corresponding ordered samples. For further details we refer to \cite{frufru:2011}.

\section{Simulation study}
\label{simulation}
We designed a simulation study to evaluate the performance of the proposed model, with the aim of assessing the performance of our approach in recovering the latent coordinates, the number of clusters, and the cluster allocation. 

We define four different simulation scenarios, in order to analyze the performance of the estimation procedure under different levels of complexity of the latent space and of the observed multidimensional network. In each scenario, we consider a bi-dimensional latent space with latent positions generated from a mixture of $G = \{2, 3, 4\}$ components. Two alternative settings for the size of the multiplex are considered: $(n=25, K=3)$ and $(n=50, K=5)$. For each scenario and setting, $10$ datasets are generated. 

The first three scenarios are constructed by generating the latent space according to proposed model specification presented in Section \ref{modello:mio}, with scenarios that differ with respect to the cluster sizes and the overlapping between the clusters. The fourth scenario is constructed to analyze the results of the estimation procedure when the model is incorrectly specified. More details about the data generation are reported in Appendix~\ref{appendix:sim}. In general, the four scenarios are structured as follows: 
\begin{itemize}
    \item Scenario I. The latent coordinates have been generated according to the proposed ILPCM and the cluster are of approximately equal size. Mixture component memberships are generated from a Multinomial distribution with probabilities $\pi = (0.5,0.5)$ for the case $G=2$, $\pi = (1/3, 1/3, 1/3)$ for $G = 3$, and $\pi = (0.25,0.25,0.25,0.25)$ when G = 4. Simulated component parameters are generated such that the clusters are sufficiently separated, resulting in probability of connections for nodes in the same cluster taking values in the interval $[0.40,0.80]$, while values are, on average, smaller than 0.20 when the two nodes do not belong to the same cluster. This scenario corresponds to cohesive and equally sized groups. 
    
    \item Scenario II. Here, latent coordinates are simulated as in Scenario I, with mixture components so that the average edge probabilities take values similar to those of Scenario I for nodes in the same or in different clusters. Differently from Scenario I, most of the nodes are assigned to a single big component, while the rest is spread into smaller clusters. We generate cluster sizes from a Multinomial distribution with probability parameters $(0.2,0.8)$, $(0.1, 0.1, 0.8)$, and $(0.1,0.1,0.1,0.7)$ for the cases of $G=2, G = 3$, and $G = 4$ respectively. This scenario corresponds to the situation where the majority of the nodes of the multiplex belong to a large cohesive group, while few other groups of actors are isolated and rarely interact with the rest of the nodes.
    
     \item Scenario III. The aim of this scenario is to evaluate the ability of the estimation procedure in recovering the clustering structure when the interactions among the clusters are frequent. Differently from scenarios I and II, the mixture component parameters are generated to have higher variability and overlapping between the clusters (see Appendix~\ref{appendix:sim}). This results in nodes belonging to different clusters having a maximum edge probability equal to 0.40 on average (instead of 0.20 in Scenario I). The increased probability of connection between clusters results in more dense network views and communities that are more difficult to separate. In this scenario clusters are of different sizes, and the data have been simulated according to the ILPCM model for $G={2, 3, 4}$ and mixing weights simulated from a Uniform distribution in the interval $[0.3, 0.8]$ and then normalized.
     
     \item Scenario IV. This last scenario is particularly complex to face as it considers a misspecified model for the distribution of the latent coordinates. The aim is to evaluate up to what extent latent coordinates and cluster allocations can still be recovered by the proposed estimation procedure.
     We simulate the latent coordinates using a mixture of multivariate non-central Student $t$ distributions, with $3$ degrees of freedom. The component sizes are generated as in Scenario I. We generate mixture component parameters such that the average edge probability values for pairs of nodes belonging to the same cluster are in the interval $[0.15, 0.5]$, while the average probability of a link is 0.3 when the two nodes do not belong to the same cluster. The increased variability of connections and the use of Student $t$ component distributions results in clusters that overlap significantly and have shapes substantially deviating from those assumed under the ILPCM.
     
\end{itemize}
In all scenarios we run the MCMC procedure described in Section \ref{estimation} for $60000$ iterations, with a burn-in of $10000$ iterations. We set the dimension of the latent space $p$ equal to $2$, is the main intent is to visualize the multiplex and the clustering. Following \cite{mio}, we opt for the following choices of hyperparameters: $m_{\alpha} = m_{\beta} = 0$, $\nu_{\alpha} = \nu_{\beta} = 3$, $\nu_{1} = n$, $\nu_2 =1$, $\tau_{\alpha} = \tau_{\beta} = \tau_z =1$. Small variations of these hyperparameter values have also been tested, but they did not affect substantially the simulation results. The hyperparameters of the prior distribution on the $\psi$ parameter are set to $\xi_1 = 1$ and $\xi_2 =2$ \citep{bernardo:1988, grazian:2015}.

\subsection{Simulation results}
As our interest lies in recovering the latent space positions and the clustering structure, we will primarily focus on such aspects. However, we briefly mention that the actual intercept values $\ak$ are always included within the $99\%$ credible intervals, while the scale coefficients $\bk$ tend to be overestimated, due to underestimation of the latent distances. Nonetheless, the products between the scale coefficients and the distance $\bk d_{ij}$ are well recovered for all nodes and views.  

\begin{figure}[t]%
    \centering
    {{\includegraphics[scale=0.52]{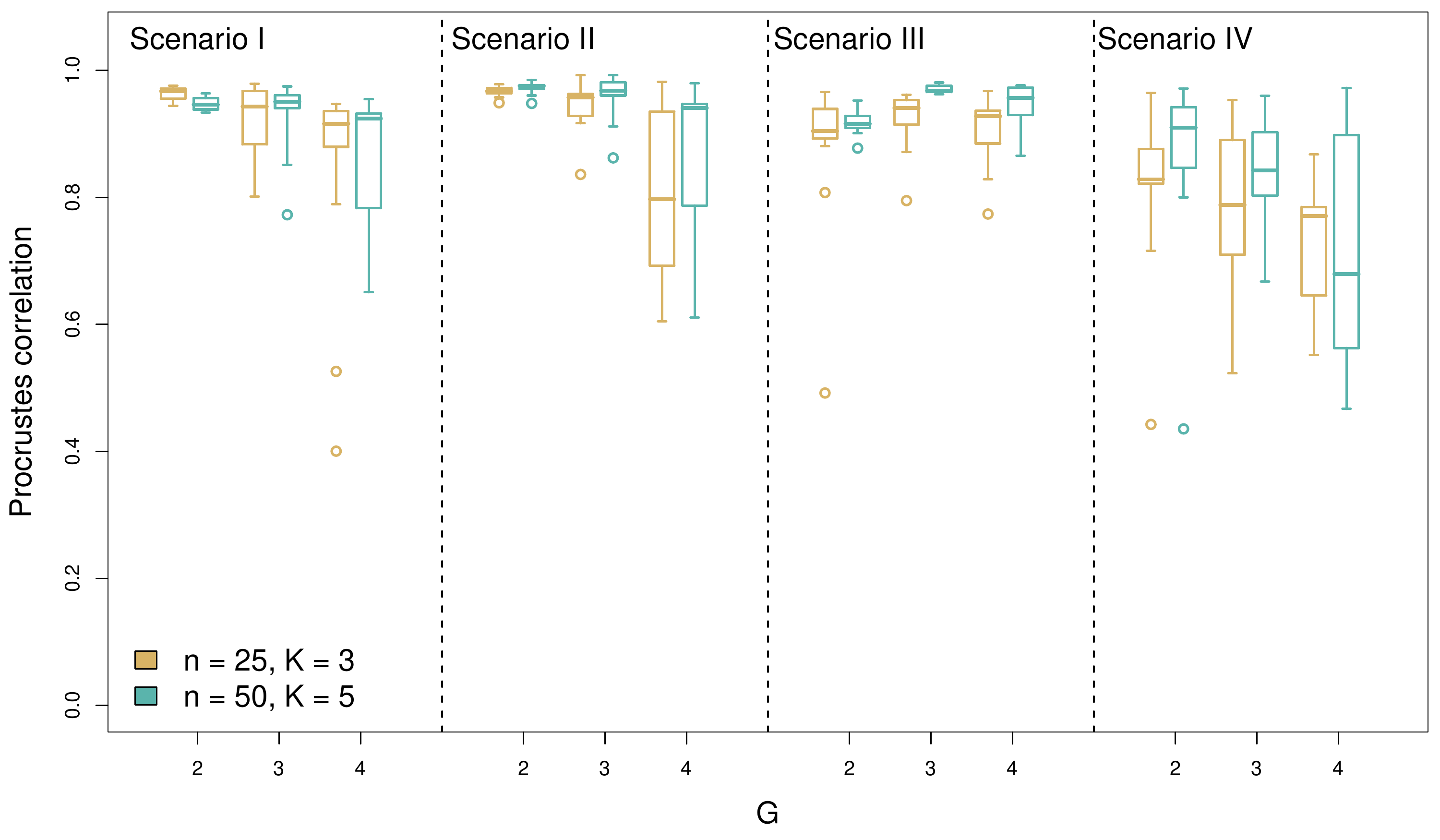} }}%
    \caption{Simulation study. Procrustes correlation values between the estimated latent positions and the simulated ones, conditioning on the four scenarios, the data dimensionality and the number of components.}%
    \label{fig:sim_prot}%
\end{figure}

Figure \ref{fig:sim_prot} reports the values of the Procrustes correlation between estimated (posterior means) and simulated latent coordinates, for the different scenarios.  The values are quite large, independently of the scenario considered. This suggests that the proposed estimation method is able to recover quite well the latent distances between the nodes, even when the mixture components are not Gaussian (Scenario IV).

\begin{figure}[t]%
    \centering
    {{\includegraphics[scale=0.52]{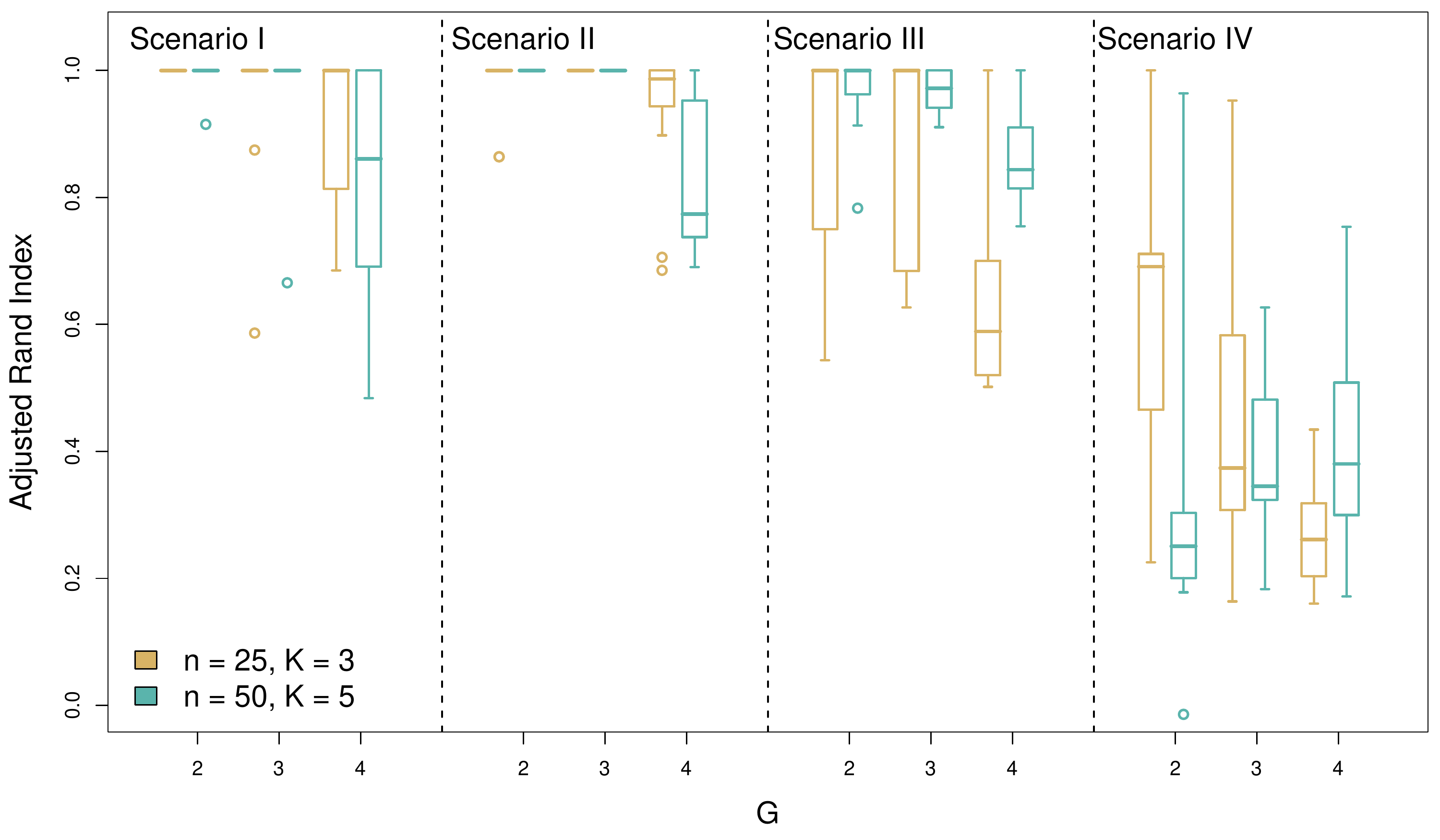} }}%
    \caption{Simulation study. Adjusted Rand Index values between the estimated labels and the simulated ones, conditioning on the four scenarios, the data dimensionality and the number of components.}%
    \label{fig:sim_ari}%
\end{figure}

Figure \ref{fig:sim_ari} reports the values of the Adjusted Rand index (ARI) \citep{Hubert:1985} computed between the actual and the estimated cluster labels. Values of this index larger than $0.6$ are usually regarded of good quality in the literature, with $1$ corresponding to a perfect match between the estimated and the actual clustering of the nodes.
In Scenarios I, II, and III, the median ARI value is generally above $0.6$, and cluster assignments are almost perfectly recovered when $G=2$ and $G=3$, suggesting that overall the method is able to reconstruct the latent space and infer the cluster membership of the nodes. ARI values tend to be lower when $G=4$, an indication that the number of cluster might not be correctly identified in this situation, and likely underestimated. However, classification performance in terms of ARI is still good. The median ARI values in Scenario IV are lower than those in the other scenarios, and they range between $0.2$ and $0.7$. This is expected, since the model is incorrectly specified and there is significant overlapping, and Gaussian component distributions do not correctly model clusters generated from Student $t$ distributions. With this in mind, an ARI larger than $0.20$ still denotes a clustering solution of acceptable quality. 

\begin{figure}[t]%
    \centering
  {{\includegraphics[scale =.42]{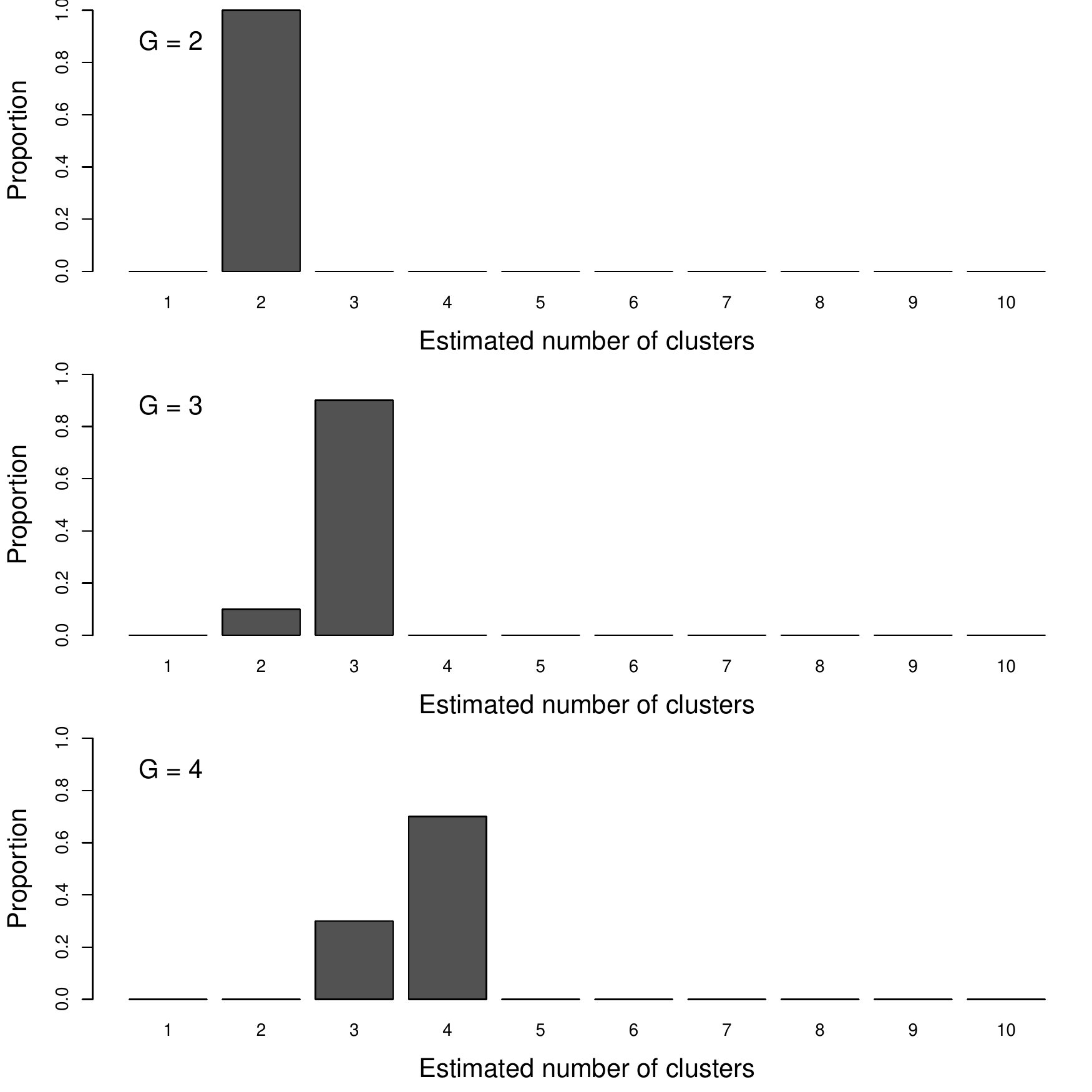}} }
  {{\includegraphics[scale =.42]{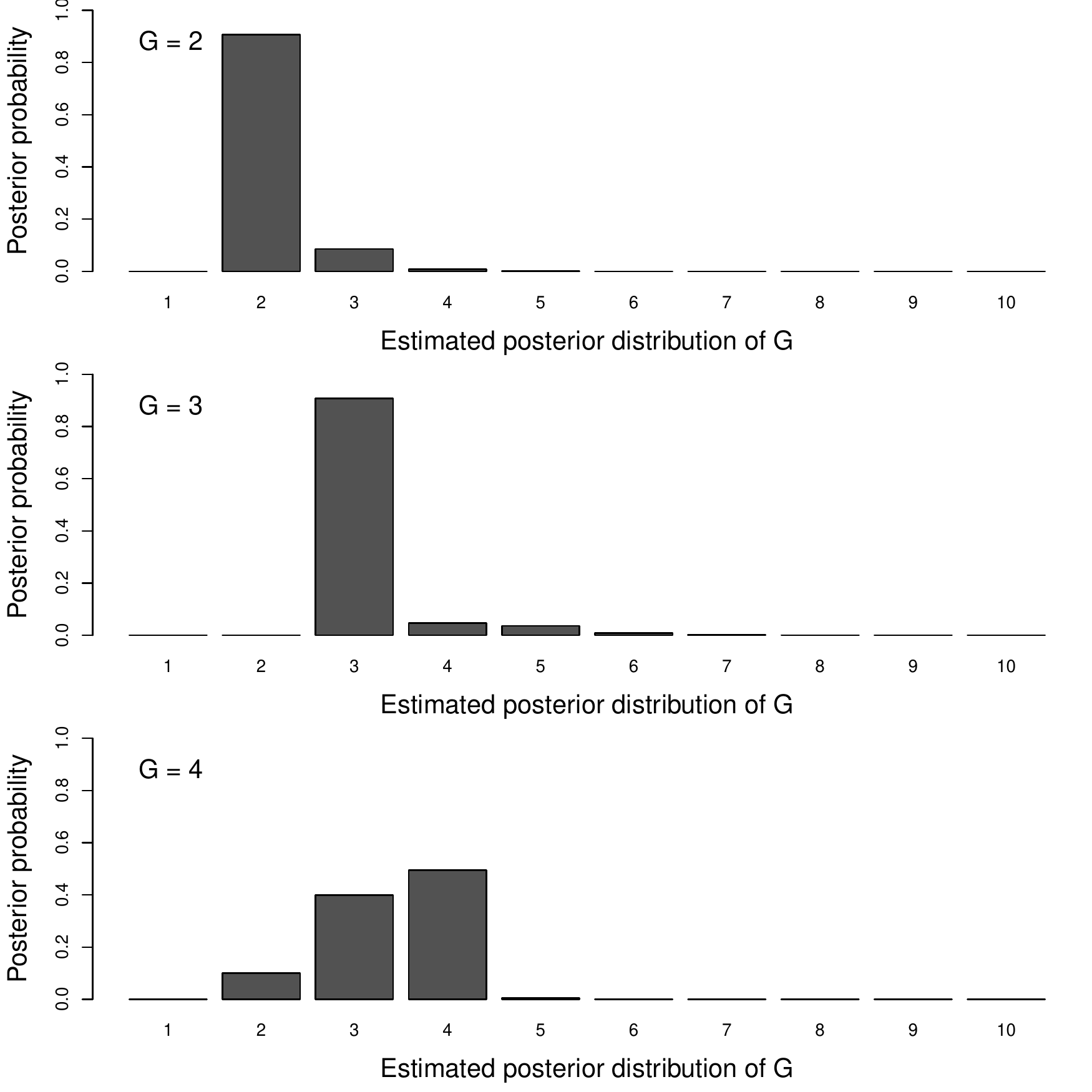} }}
        \caption{Simulation study Scenario I. Estimated number of mixture components, for $n=25$ and $K=3$ (left plot) and $n=50$ and $K=5$ (right plot).}
        \label{fig:simS1}%
\end{figure}

\begin{figure}[t]%
    \centering
  {{\includegraphics[scale =.42]{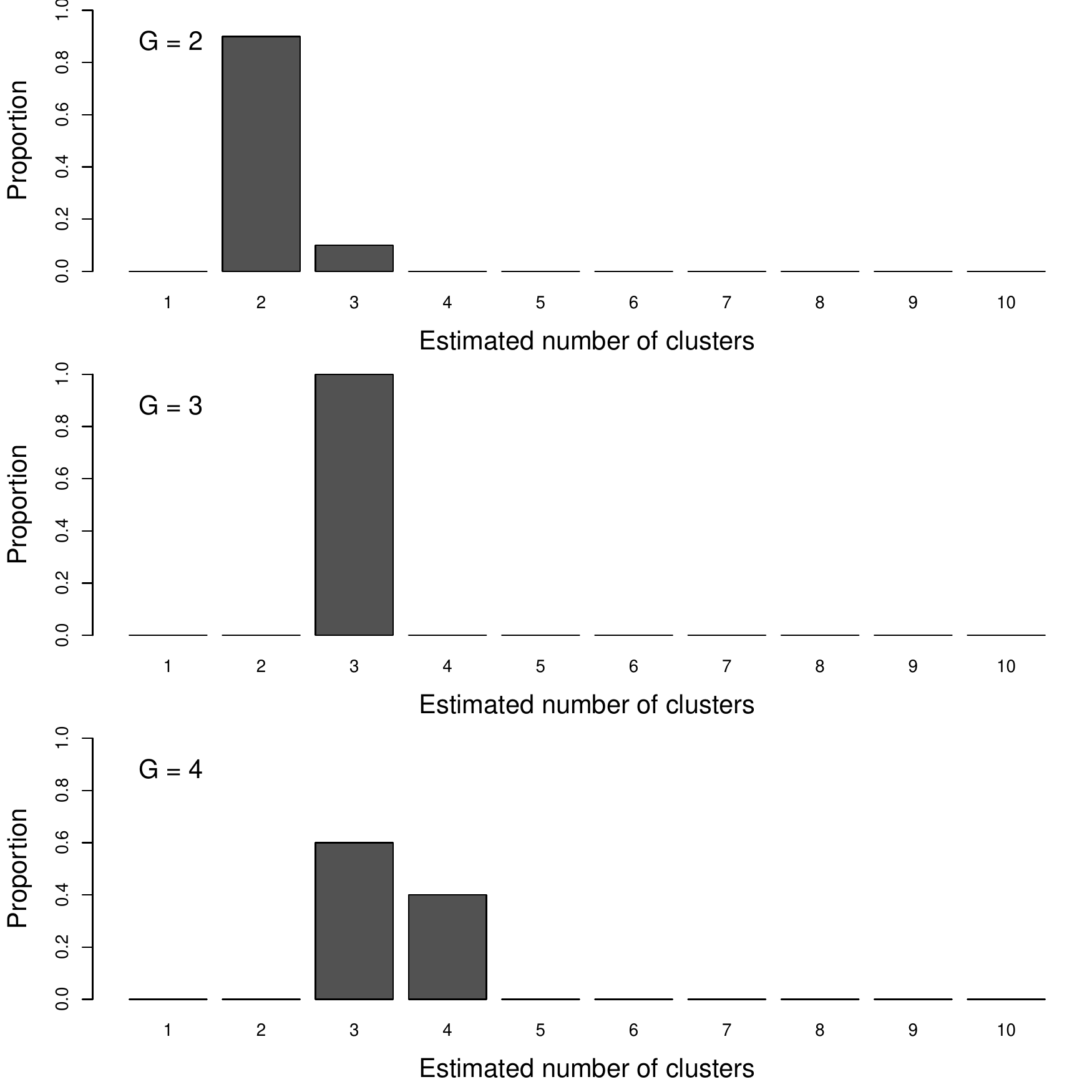}} }
  {{\includegraphics[scale =.42]{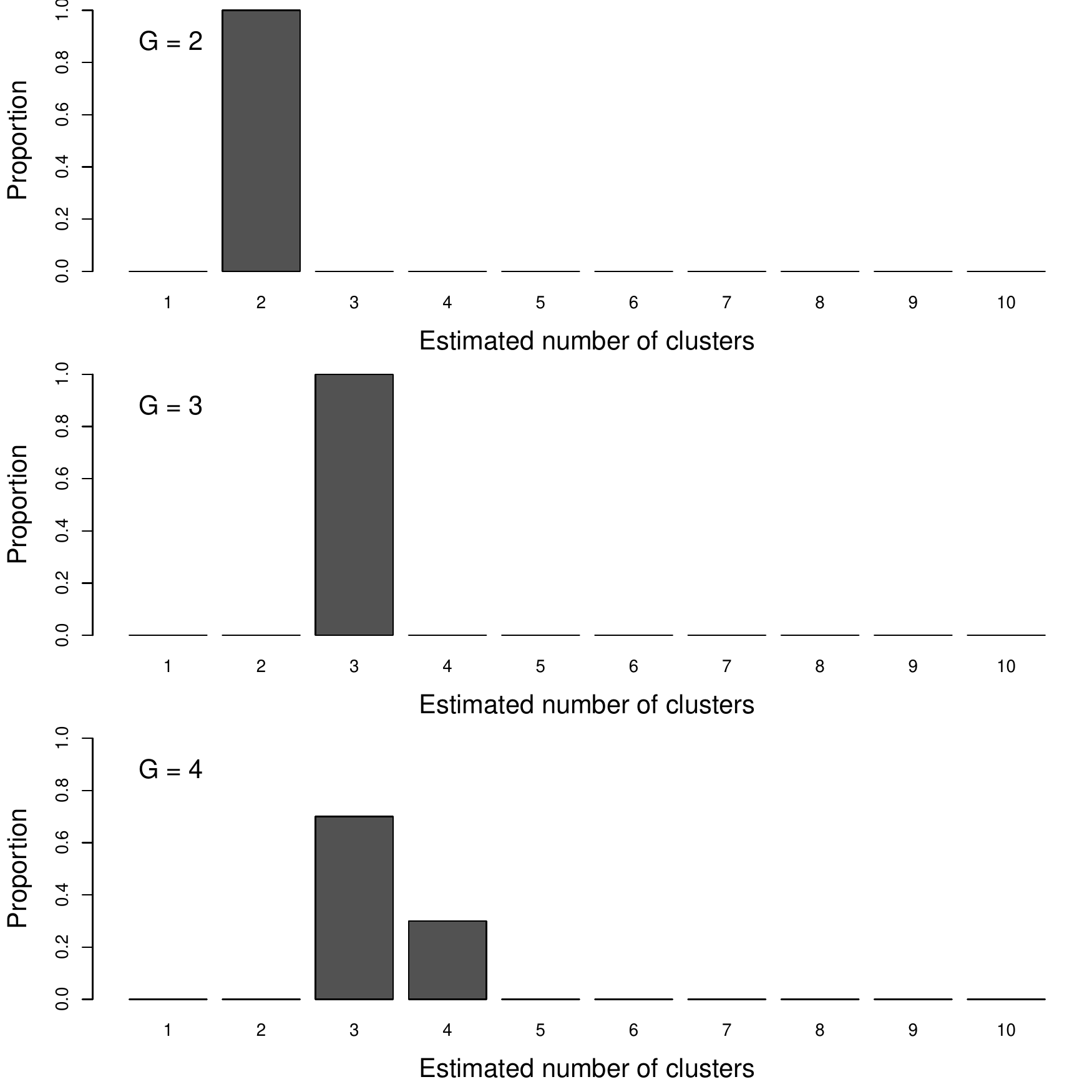} }}
        \caption{Simulation study Scenario II. Estimated number of mixture components, for $n=25$ and $K=3$ (left plot) and $n=50$ and $K=5$ (right plot).}
        \label{fig:simS2}%
\end{figure}

\begin{figure}[t]%
    \centering
  {{\includegraphics[scale =.42]{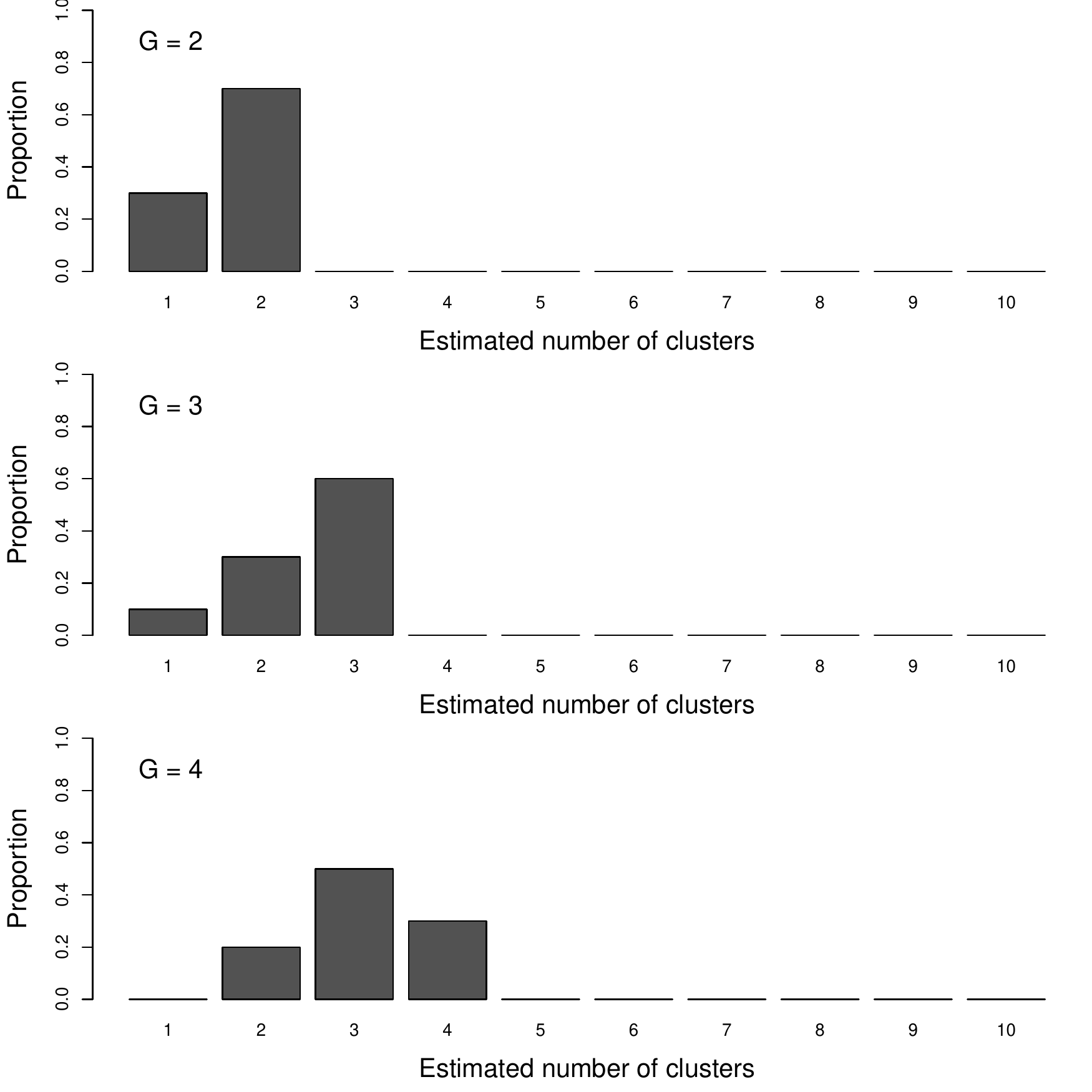}} }
  {{\includegraphics[scale =.42]{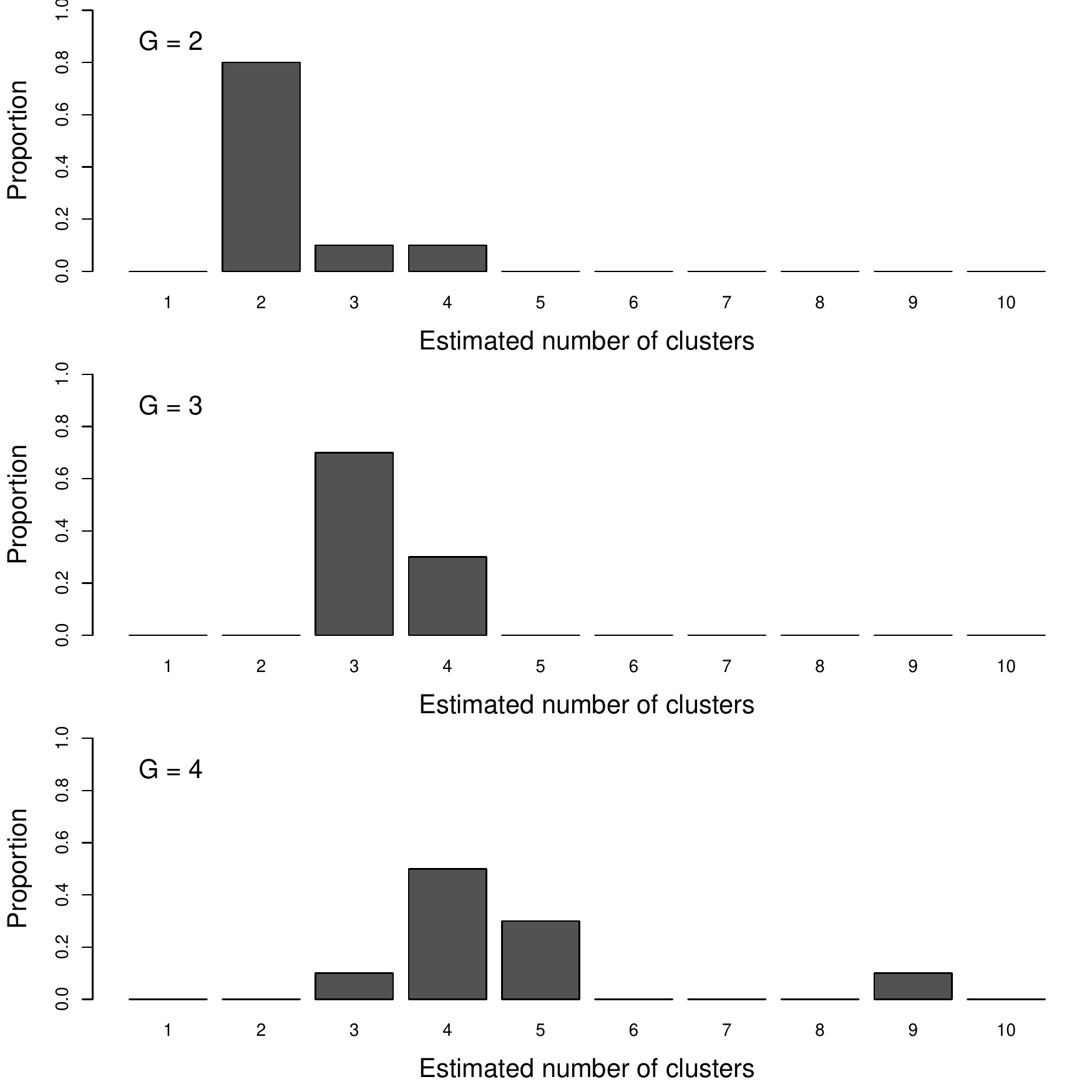} }}
        \caption{Simulation study Scenario III. Estimated number of mixture components, for $n=25$ and $K=3$ (left plot) and $n=50$ and $K=5$ (right plot).}
        \label{fig:simS3}%
\end{figure}

\begin{figure}[t]%
    \centering
  {{\includegraphics[scale =.42]{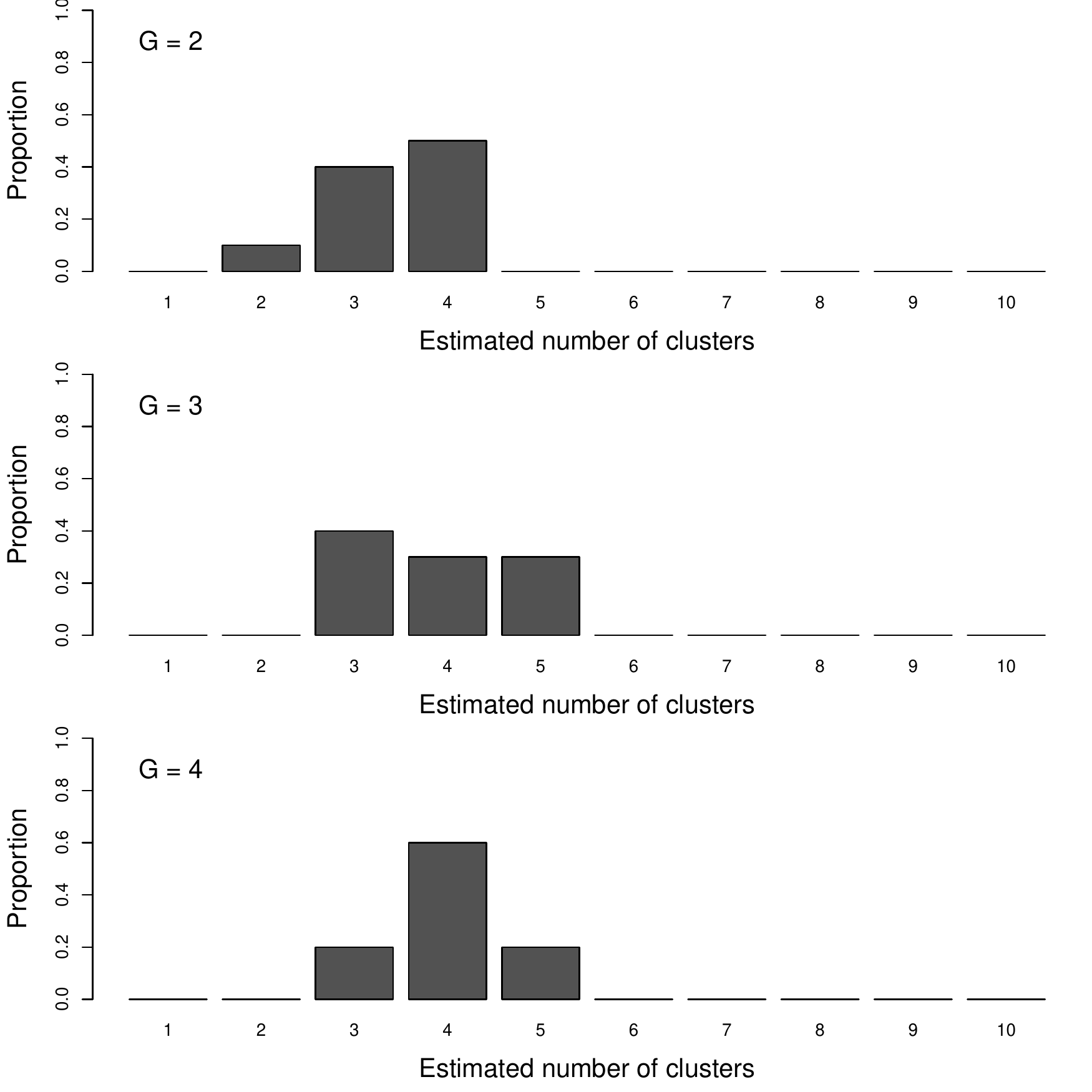}} }
  {{\includegraphics[scale =.42]{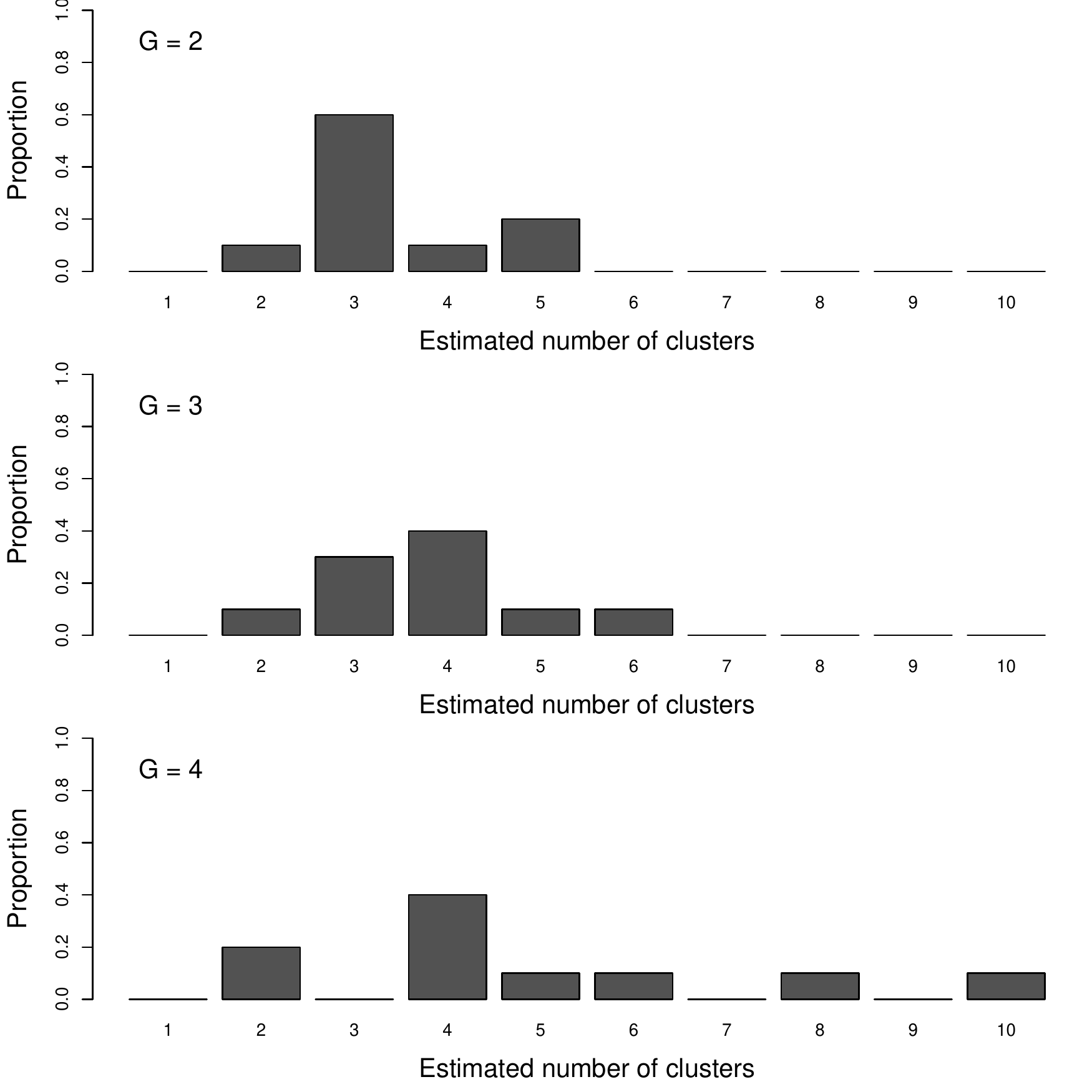} }}
        \caption{Simulation study Scenario IV. Estimated number of mixture components, for $n=25$ and $K=3$ (left plot) and $n=50$ and $K=5$ (right plot).}
        \label{fig:simS4}%
\end{figure}

Figures \ref{fig:simS1}-\ref{fig:simS4} report the frequency distribution of the estimated number of clusters in the four simulation scenarios. 
In the first three scenarios, the actual number of clusters tends to be generally well recovered. The only exception is when $G=4$ under Scenarios II and III, where the number of components tends to be slightly underestimated, and often solutions with $3$ clusters are preferred. This is a reasonable result, considering that the number of nodes is quite small in comparison with the number of components, that some clusters contain only few nodes (Scenario II), and that there is significant overlapping and frequent interactions between clusters (Scenario III). Scenario IV depicts a very difficult situation, as the model is misspecified and there is overlapping among the clusters. Generally, the number of clusters is overestimated and this appears to be caused by the fact that a larger number of Gaussian components are required to approximate a mixture of Student t distributions. 

Overall, results are satisfactory, as the positions of the nodes in the latent space are consistently well recovered under all scenarios, and the classification performances are in line with the complexity of the clustering problem.
\clearpage

\clearpage
\section{Applications}
\label{data}
\subsection{Lazega lawyers}
\label{lazega}
The Lazega lawyers data are a collection of $K=3$ networks, describing different types of relations between $N=71$ attorneys in the ``SG\&R'' law firm, collected in New England between 1988 and 1991 \citep{Lazega:2001}. A number of node related covariates is also recorded: seniority, formal status, sex, office in which lawyers work, age, lawschool attended, and the type of practice on which each node work.
More in details, the views of the multiplex record exchange of advice (observed density $0.18$), friendship (observed density $0.12$), and co-working (observed density $0.15$) interactions between the lawyers. 

\begin{figure}[t]
    \centering
  {{\includegraphics[scale =.4]{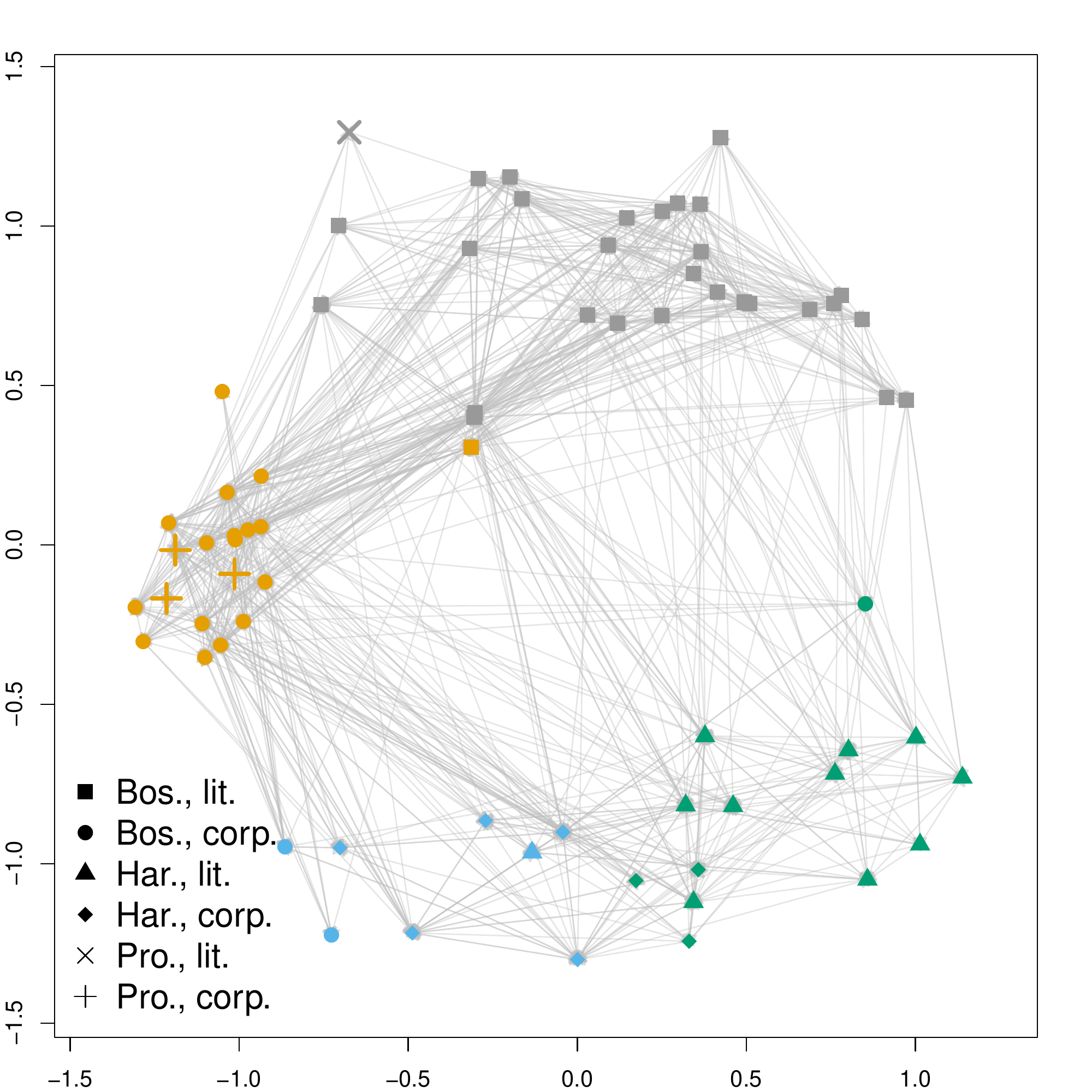}} }
  {{\includegraphics[scale =.4]{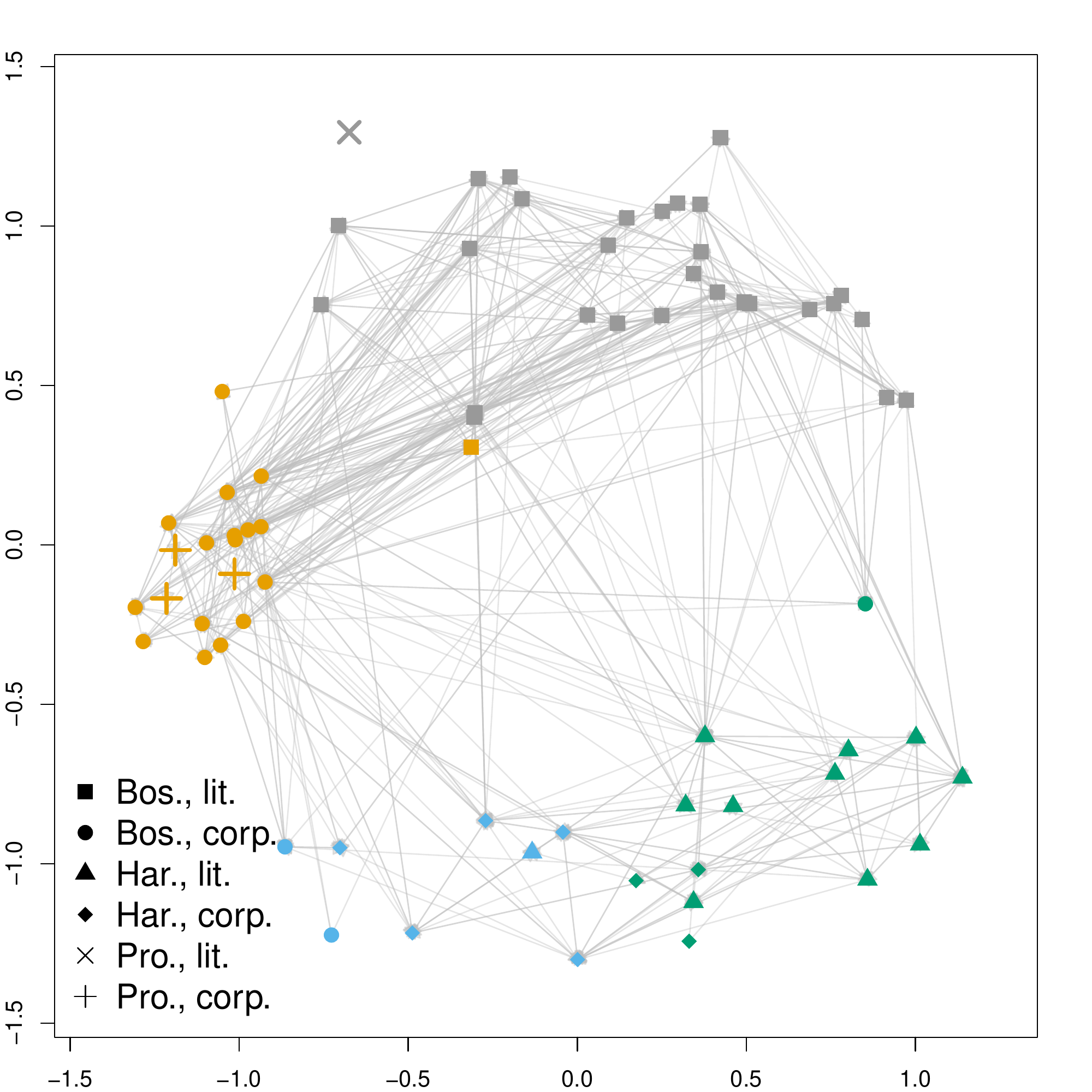} }}
  {{\includegraphics[scale =.4]{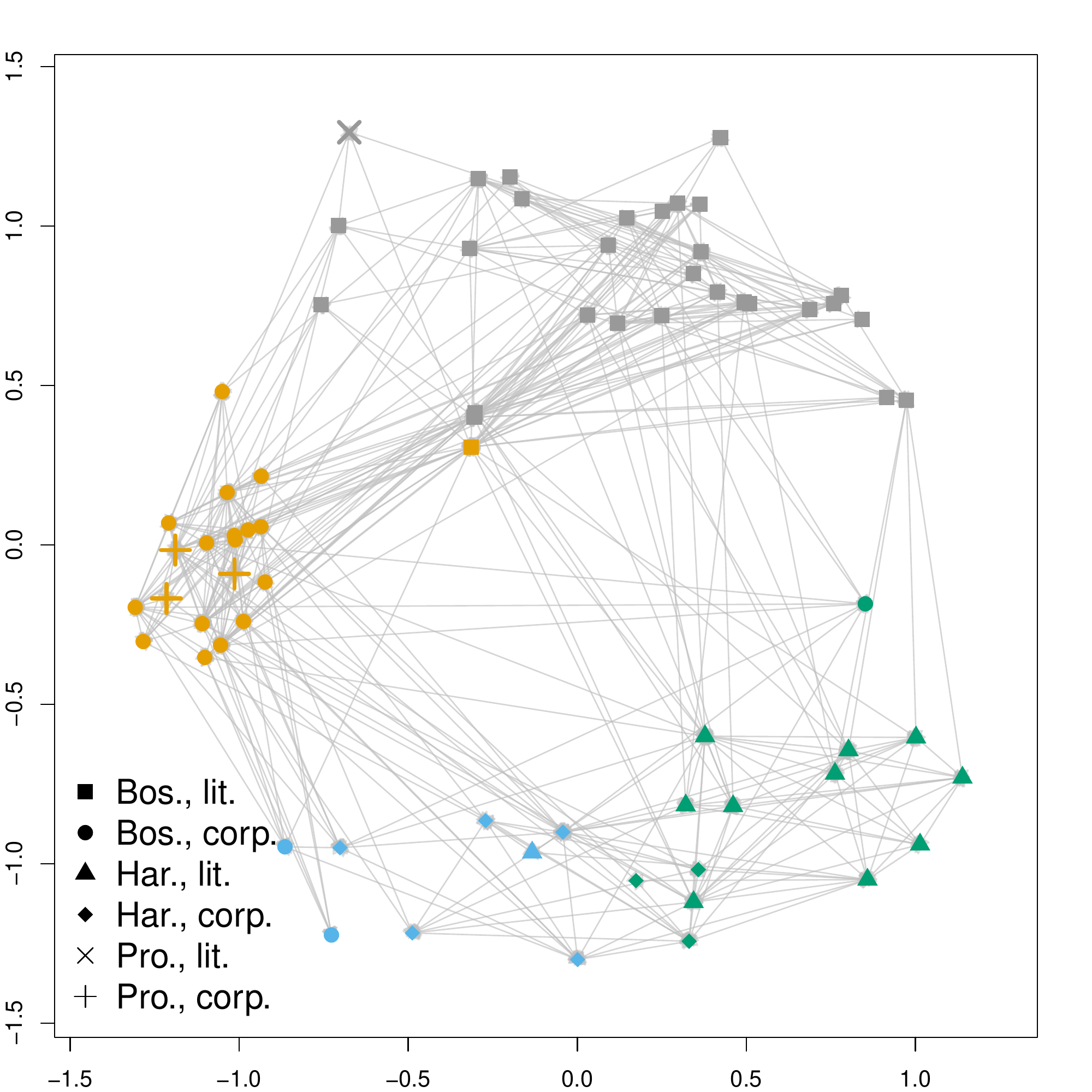} }}
        \caption{Lazega lawyers data. Estimated latent space and observed connections in the advice (top left), friendship (top right) and work (bottom) networks. Point types represents the combinations of workplace (Boston, Hartford and Providence) and type of practice (litigation and corporate), while colours represent estimated clusters.  }
           \label{fig:lazega}%
\end{figure}

We have run the MCMC procedure described in Section \ref{estimation} for $60000$ iterations, with a burn-in of $10000$ iterations. The hyperparameters have been set as in the simulation study, see Section \ref{simulation}. For visualization purposes, the dimension of the latent space has been set to $p=2$.
The cluster allocation posterior distribution was post-processed using the method by \cite{wade:2018}, described in Section \ref{postproc}, and $G=4$ clusters were retrieved. 
The corresponding estimated partition is displayed in Figure \ref{fig:lazega}, where estimated latent coordinates are shown and coloured according to the estimated clustering. Connections between the lawyers in the different networks are represented as grey arrows joining pairs of latent coordinates. In addition, the information about the office and the type of practice a lawyer works on is represented by the type of point. Inferred allocations are in good agreement with the partition defined by these two covariates, with an Adjusted Rand Index of $0.74$.
The largest cluster is the grey one, grouping together $29$ litigation lawyers, representing almost all litigation attorneys from the Boston office ($28$ out of $29$) and one from the Providence office.
The second largest component is the yellow one, where $19$ corporate lawyers are grouped, $16$ from the Boston office and $3$ from the Providence office. An extra lawyer from the Boston office is allocated to this group, although working on litigation practices. The green component is formed by $14$ lawyers, mainly working on litigation practices, $13$ from the Hartford office and $1$ from the Boston office. Last, the smallest component (blue) includes $8$ attorneys, the majority of whom are corporate lawyers, $6$ from the Hartford office and $2$ from the Boston one. In general, the obtained clustering and the office and practice covariates seem to suggest a reasonable degree of separation between lawyers working on different practices. Also, it is suggested that while the Hartford and Providence offices tend to collaborate with the Boston office, they never interact among themselves.

\subsection{Aarhus Computer Science department}
\label{Aarhus}
The Aarhus Computer Science department is a multiplex composed of $5$ types of interaction between $N=61$ members of the department of Computer Science at Aarhus University. The nodes are characterized by their role in the department: admin, PhD student, Postdoctoral researcher, Assistant Professor, Associate Professor, Full Professor, and Emeritus Professor. For more information on the data, we refer to \cite{Magnani:2013}. Views collect different interactions between the members of the Computer Science department: work together, lunch together, being friends on Facebook, participating to leisure activities together, and being co-author in a scientific manuscript.

\begin{figure}[t]%
    \centering
  {{\includegraphics[scale =.4]{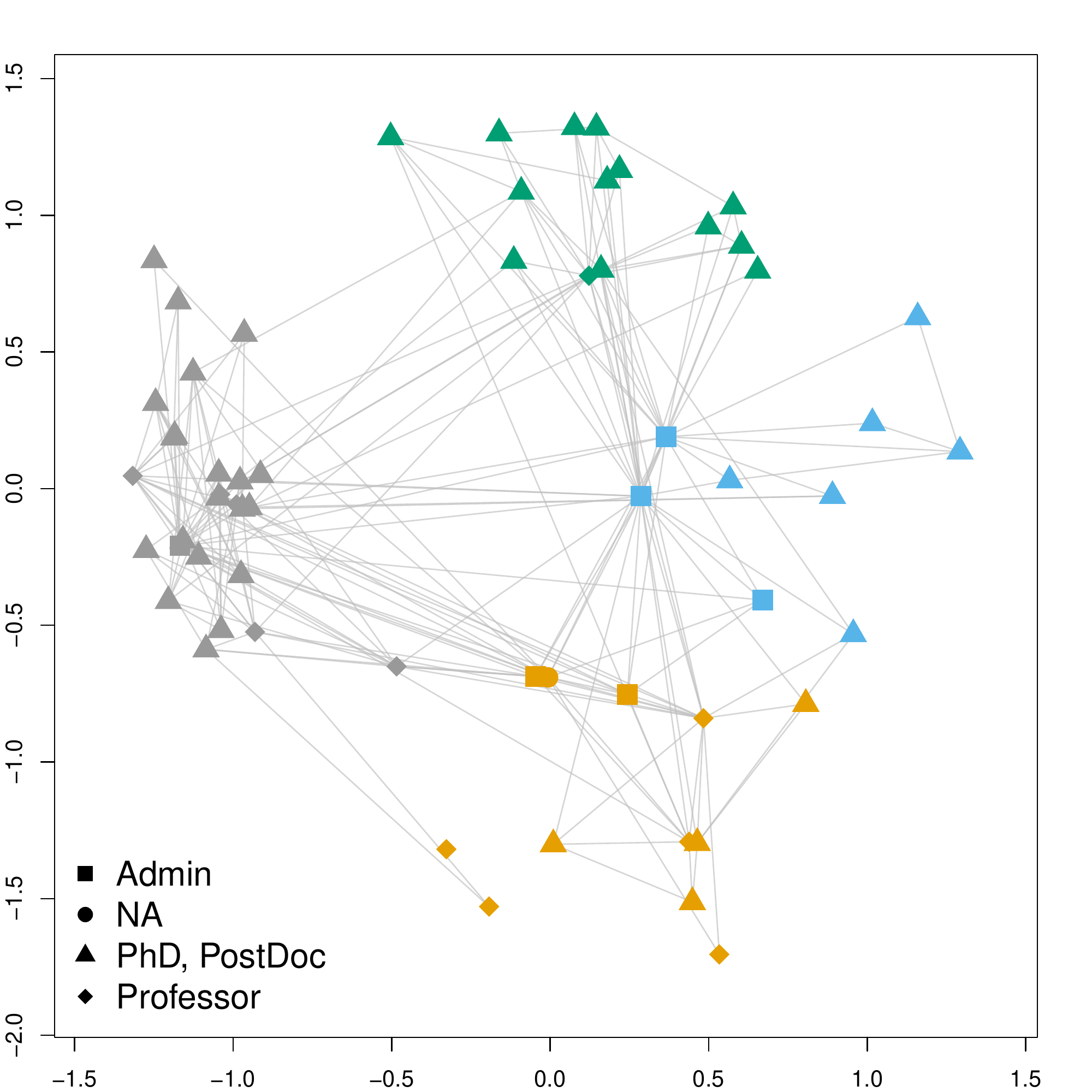}} }
  {{\includegraphics[scale =.4]{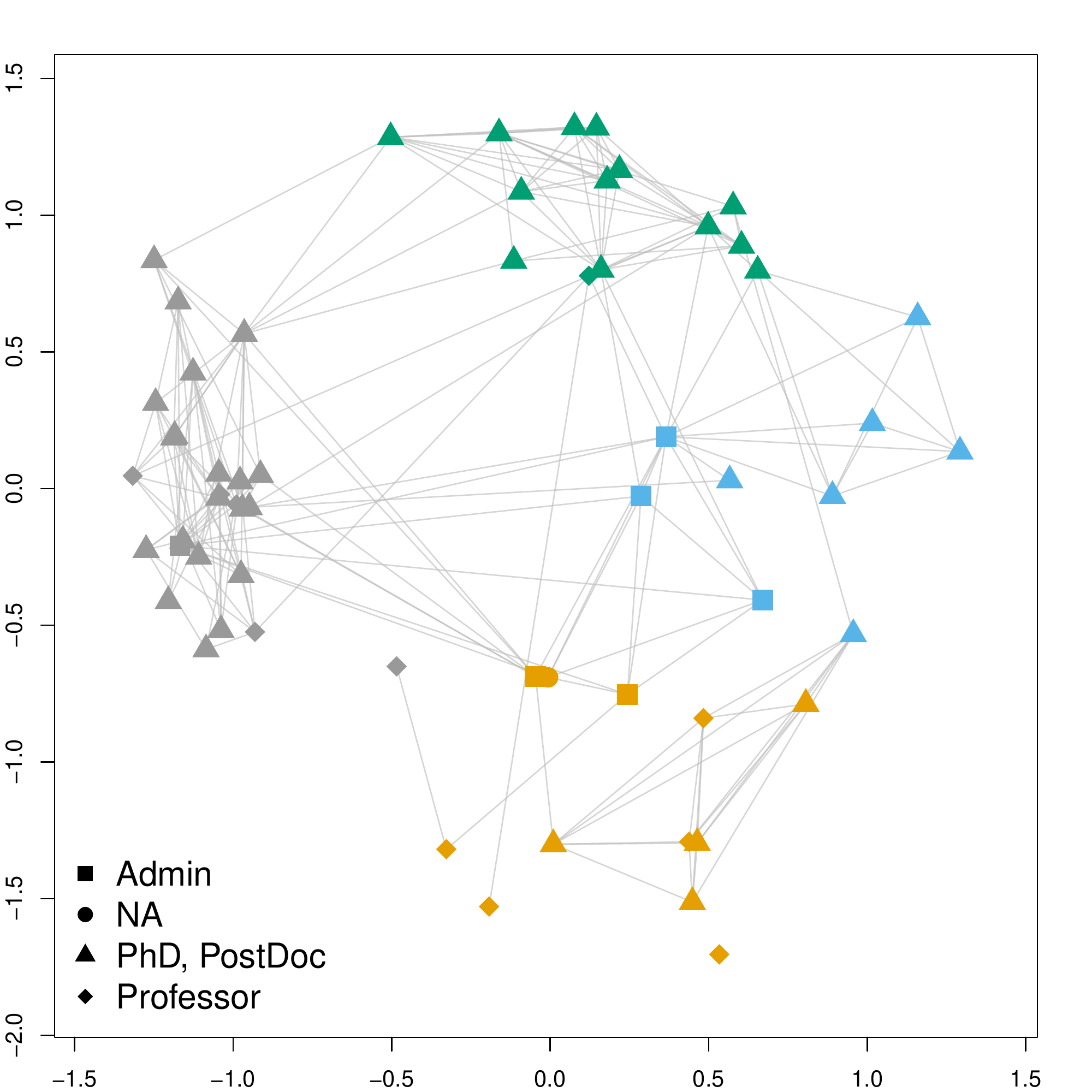} }}
  {{\includegraphics[scale =.4]{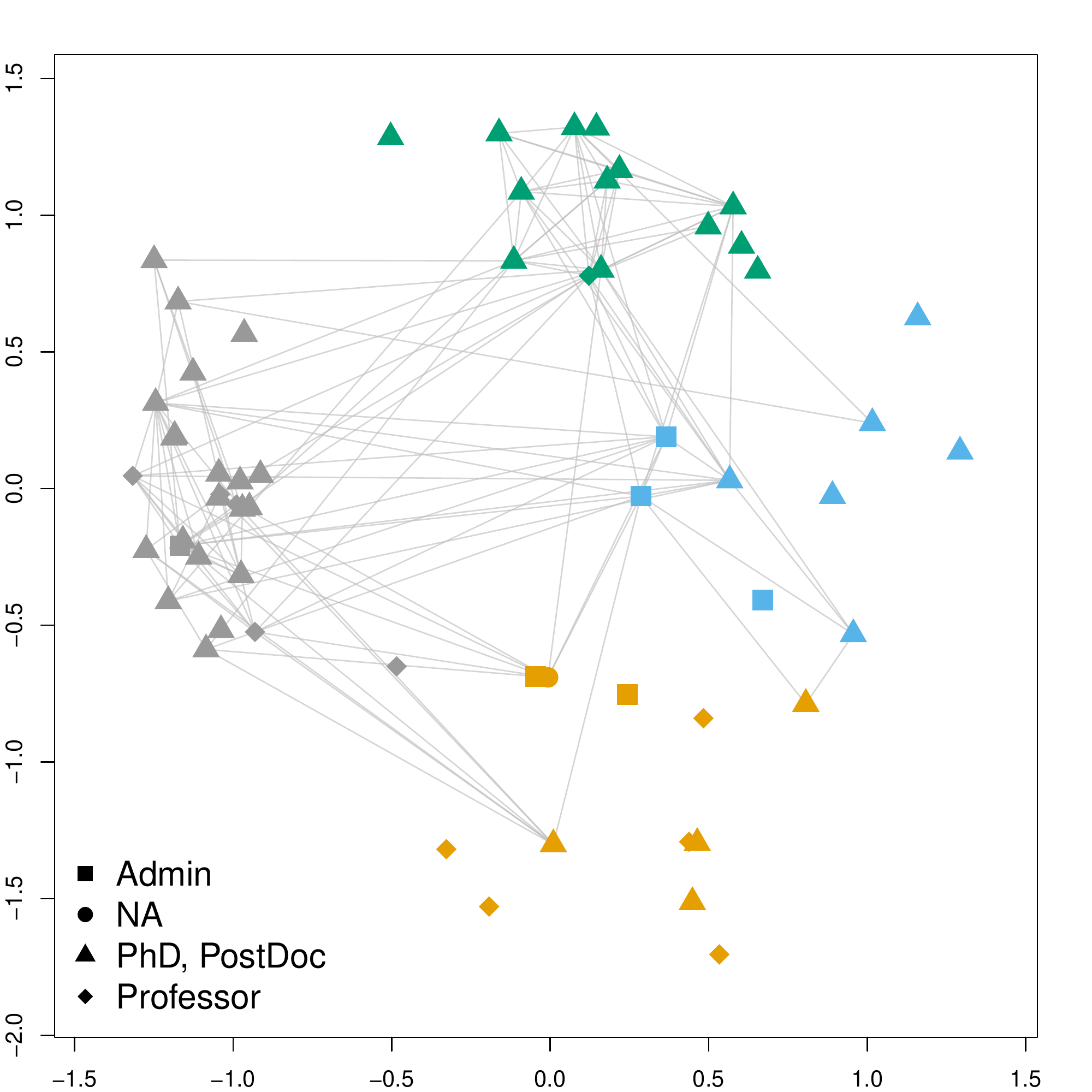} }}
  {{\includegraphics[scale =.4]{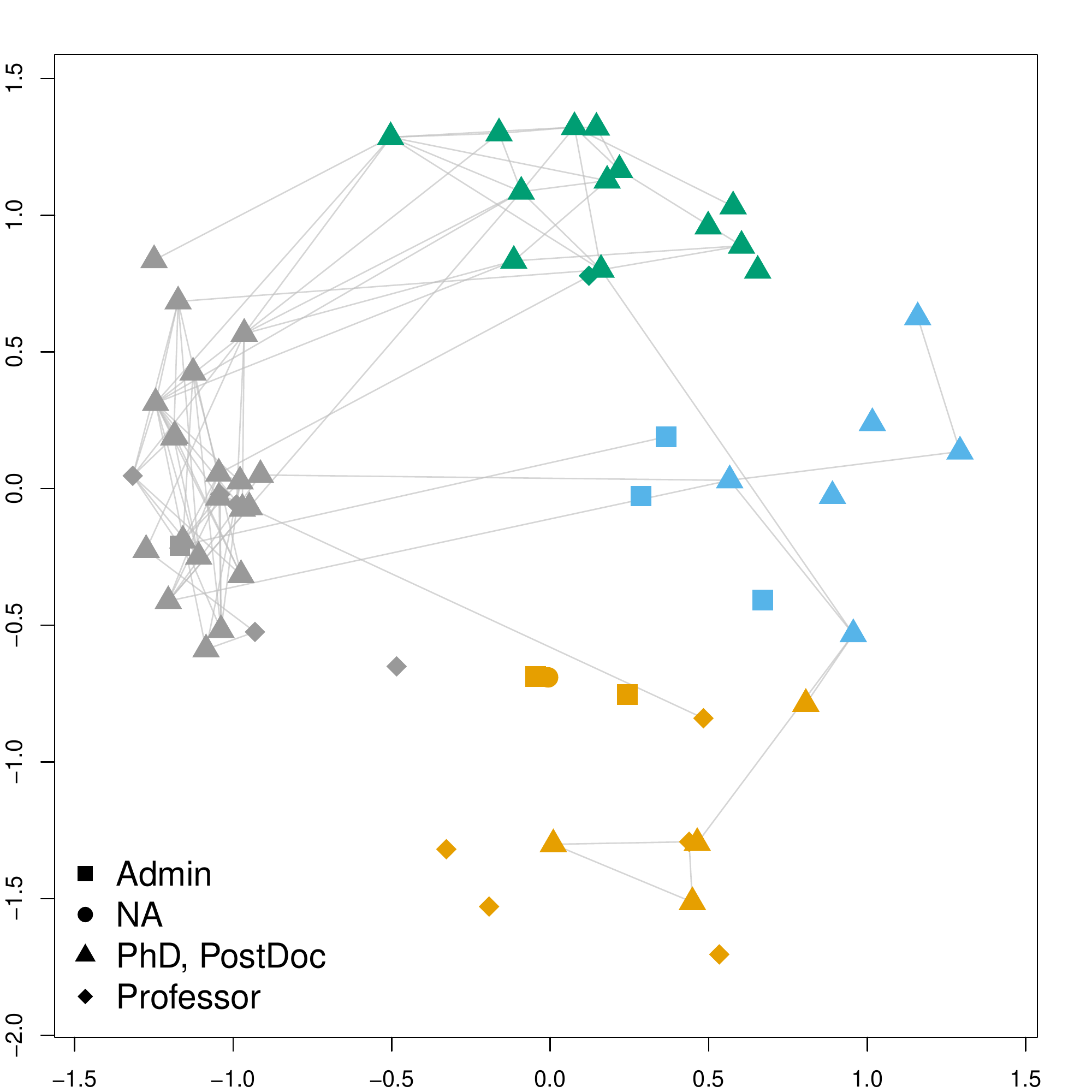} }}
  {{\includegraphics[scale =.4]{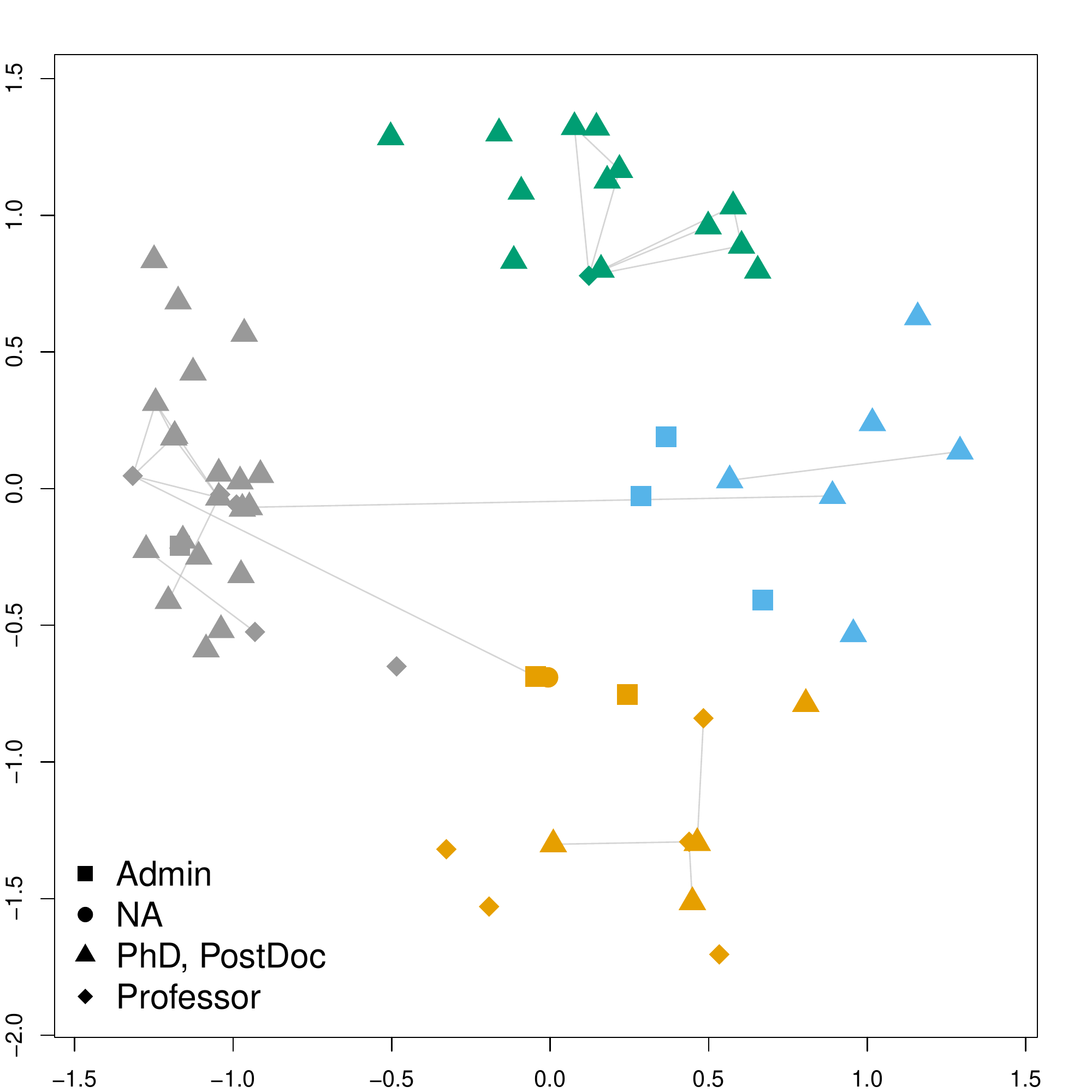} }}
        \caption{Aarhus Computer Science department data. Estimated latent space and observed connections in the work (top left), lunch (top right), facebook (middle left), leisure (middle right) and   coauthor (bottom) networks. Point types represents the position of the node: Admin, Unknown (NA), PhD or Postdoc, or Professor (Assistant, Associate, Full or Emeritus). }
           \label{fig:aarhus}%
\end{figure}

We have run the proposed MCMC for $60000$ iterations ($10000$ iterations burn-in). Similarly, the hyperparameters have been set as in the simulation study, and the dimension of the latent space has been set to $p=2$.
To estimate the clustering of the nodes we used the method of Section \ref{postproc}, which suggests $G=4$ clusters. 
Figure \ref{fig:aarhus} reports the estimated latent space, with the latent coordinates connected by an arrow if they are linked in a given view of the multiplex. The number of nodes allocated per cluster is $26$, $12$, $9$ and $14$, respectively for the grey, yellow, blue and green component. 
In addition, the role of a member of the department is represented with different symbols. Squares indicate admin, triangles indicate PhD students and PostDocs, and diamonds indicate Professors (Assistant, Associate, Full, and Emeritus). There is a further circle-shaped point denoting a member of the department whose role is unknown. Admin nodes are quite central in the latent space, and indeed are involved in a large number of interactions in the work and lunch networks, denoting their vital role in the everyday life of the department. PhDs and PostDocs are the vast majority of the nodes. They are split across the four clusters, and tend to have quite cohesive interactions within the cluster they have been assigned to. This is quite evident in the co-authorship network, which is quite sparse, and where all interactions except two happen within the components. 

\clearpage
\section{Discussion}
\label{discussion}
In this work we introduced the infinite latent position cluster model (ILPCM) for clustering the nodes of multidimensional social network data.
The proposed ILPCM allows to address transitivity and clustering in multiplexes via a latent space representation of the nodes, whose coordinates are assumed to be drawn from a mixture of Gaussian distributions. Differently from existing clustering methods for network data, we exploit a Bayesian nonparametric approach and postulate that the number of mixture components is potentially infinite. The resulting framework enables for joint inference on the latent social space and the clustering of the nodes, allowing a parsimonious representation of the otherwise complex multiplex object.
Model inference is carried out within a Bayesian hierarchical approach via a MCMC algorithm, whose performance has been tested in a simulation study, showing good quality results.
The ILPCM bypasses the model choice issue of selecting the number of components, avoiding the computational burden of fitting and comparing multiple models.
The proposed method has been applied to two social multidimensional networks, concerning professional and personal relations among colleagues working in the same institution. 
For both multiplexes, the ILPCM has been useful in capturing and visualizing clustering structures between the nodes in a lower dimensional space.

In all examples, we set the latent space dimension equal to $2$, which allowed to visualize the multidimensional networks in a clear and simple way. Assuming a two-dimensional latent space is usual practice when modelling network data via latent space models. However, selection of the appropriate number of latent dimensions can be a crucial task when the focus is other than simple visualization. The problem is still open in the literature, with only few solutions explored \citep[][for example]{durante}, and it may constitute further extension of the proposed ILPCM. 

The code implementing the proposed ILPCM is currently  available upon request and will be incorporated shortly in the R package \texttt{spaceNet} \citep{spaceNet}, available on CRAN.

\bibliographystyle{agsm}
\bibliography{Bibliografia}

\clearpage
\appendix

\section{Appendix: Proposal and full conditional distributions}
\label{appendice3}
This appendix presents the full conditional and proposal distributions employed for inference on the latent space in the MCMC algorithm described in Section \ref{estimation}.

\subsection{Latent coordinates}
\label{appendix:latent_coordinates}
For ease of notation, we omit iteration indexes $t= 1, \dots, T$. To derive a proposal distribution for the latent coordinate $\mathbf{z}_i$, $i=1,\dots,n$, we fix the number of components to a constant $G$, as it is the case at each iteration of the estimation procedure illustrated in Section \ref{estimation}. Then, from equations \ref{eq:prob1} and \ref{eq:prioriZ}, the log posterior distribution for the $i^{\text{th}}$ latent coordinate can be easily derived. Without loss of information, we consider such log posterior for the $r^{\text{th}}$ latent space dimension ($r=1,\dots,p$) being proportional to:
\[
\sum_{j \neq i}^n \sum_{k = 1}^K\Bigl[ \arch \bk d_{ij} - \log \bigl(1 +\exp \bigl\{ \ak -\bk d_{ij}\bigr\} \bigr) \Bigr] + \log \left[ \sum_{g=1}^G \pi_g \frac{1}{\sqrt{2 \pi \sigma_{rg}^2}}\exp \left\{ -\frac{1}{2\sigma_{rg}^2} (z_{ri} -\mu_{rg})^2   \right\}  \right].
\]
The double summation term in the above equation is the log-likelihood of the model, while the other term depends on the $i^{\text{th}}$ latent position prior distribution. If we define:
\[
a_{irg} = \pi_g \frac{1}{\sqrt{2 \pi \sigma_{rg}^2}}\exp \left\{ -\frac{1}{2\sigma_{rg}^2} (z_{ri} -\mu_{rg})^2   \right\}  = \exp(x_{irg}) \quad \Leftrightarrow \quad x_{irg} = \log (a_{irg} ), 
\]
we can rewrite the second term of the above log-posterior as an LSE (LogSumExp) function:
\[
 \log \left[ \sum_{g=1}^G \pi_g \frac{1}{\sqrt{2 \pi \sigma_{rg}^2}}\exp \left\{ -\frac{1}{2\sigma_{rg}^2} (z_{ri} -\mu_{rg})^2   \right\}  \right] = 
 \log \left[\sum_{g=1}^G  \exp( x_{rg} ) \right].
\]
This LSE function is bounded:
\[
\max_g\left\{ x_{ir1} , \dots , x_{irG}\right\} \leq \log \left[ \sum_{g=1}^G\exp(  x_{irg}  ) \right] \leq \max_g\left\{ x_{ir1} , \dots , x_{irG}\right\} +\log(G),
\]
that is 
\[
\max_g\bigl\{ \log(a_{ir1}) , \dots , \log(a_{irG}) \bigr\} \leq \log \left[ \sum_{g=1}^G a_{irg}  \right] \leq \max_g\bigl\{ \log(a_{ir1}) , \dots , \log(a_{irG})\bigr\} +\log(G).
\]
Now, note that
\begin{equation*}
  \log( a_{irg}) \propto -\frac{1}{2\sigma_{rg}^2} (z_{ri} -\mu_{rg})^2
\end{equation*}
is the log density of a Gaussian distribution with mean $\mu_{rg}$ and variance $\sigma_{rg}^2$. Note also that, at each iteration of the MCMC algorithm, $\max_g\bigl\{ \log(a_{ir1}) , \dots , \log(a_{irG})  \bigr\}$ should be found in correspondence of the $g^{\text{th}}$ component, where the coordinate $\mathbf{z}_i$ is currently allocated. Hence, we can approximate the LSE function with its lower bound, according to the above proportionality relation:
\[
 \log \left[ \sum_{g=1}^G \pi_g \frac{1}{\sqrt{2 \pi \sigma_{rg}^2}}\exp \left\{ -\frac{1}{2\sigma_{rg}^2} (z_{ri} -\mu_{rg})^2   \right\}  \right] \approx 
-\frac{1}{2\sigma_{rg}^2} (z_{ri} -\mu_{rg})^2 ,
\]
with $c_{ig} =1$ and $r=1,\dots,p$. Combining this approximation with a similar one of the log-likelihood term in the first equation (see the appendix results in \cite{mio}, for all $r = 1, \dots,p$, we derive the following proposal distribution for the latent position of a unit in the $g^{\text{th}}$ component:
\[
\tilde{\mathbf{z}}_{i} \mid \dots\thicksim {\cal N}_p(\bm{\mu}_{\tilde{\mathbf{z}}_{i}}, \bm{\Sigma}_{\tilde{\mathbf{z}}_{i}}) , 
\]
with
\[
\bm{\mu}_{\tilde{\mathbf{z}}_{i}} = \Bigl( \bm{\mu}_{1\tilde{\mathbf{z}}_{i}}, \dots, \bm{\mu}_{p\tilde{\mathbf{z}}_{i}} \Bigr), \quad
\Sigma_{\tilde{\mathbf{z}}_{i}} = \begin{bmatrix}
    \sigma_{1\tilde{\mathbf{z}}_{i}}^2   & \dots & 0 \\
    \vdots & \ddots & \vdots\\
    0     & \dots & \sigma_{p\tilde{\mathbf{z}}_{i}}^2    \\
\end{bmatrix} \mathbf{I},
\]
and
\[
\mu_{r\tilde{z}_{i}} = \sigma_{r\tilde{z}_{i}}^2 \left[ 2 \sum_{k=1}^K \bk \sum_{j \neq i } (\arch -w_{ij}^{(k)} ) z_{rj} + \mu_{rg}\right], \quad
\sigma_{r\tilde{z}_i}^2 = \left( \frac{1}{\sigma_{rg}^2} +2 \sum_{k=1}^K \bk \sum_{j \neq i }  |\arch -w_{ij}^{(k)} | \right)^{-1},
\]
with $w_{ij}^{(k)} = 1$ if $\ak -\bk d_{ij} > 0$ and $w_{ij}^{(k)} = 0$ otherwise.

\subsection{Cluster labels}
\label{appendix:labels}
Given the $i^{\text{th}}$ cluster label $\bm{c}_i$, following \cite{Bush:1996}, we can compute the posterior probability for the $i^{\text{th}}$ latent coordinate to be assigned to the $g^{\text{th}}$ component (the probability that $c_{ig} = 1)$, proportional to:
\[
P_{ig} = 
\begin{cases}
n_g^{(/i)} \mathcal{N}_p(\mathbf{z}_i \mid \bm{\mu}_g , \bm{\Sigma}_g) &\quad g = 1, \dots, G,\\
\psi \int \mathcal{N}_p(\mathbf{z}_i \mid \mu , \sigma^2) p( \mu, \sigma^2\mid \mathbf{z}^{(/i)})  d\mu d\sigma^2 &\quad g = G + 1,
\end{cases}
\]
where $G$ is the current number of components and $n_g^{(-i)}$ is the number of latent coordinates in the $g^{\text{th}}$ component, after having discarded the $i^{\text{th}}$ coordinate if present in that component. The above integral is that of a multivariate Student $t$ distribution:
\[
\int \mathcal{N}_p(\mathbf{z}_i \mid \mu , \sigma^2) p( \mu, \sigma^2\mid \mathbf{z}^{(/i)}) d\mu d\sigma^2 =  t_{2\nu_{1} } \left( 
m , \frac{\nu_{2} }{\nu_{1} }(1 + \tau_z )\mathbf{I}  \right).
\]
New cluster label vectors $\mathbf{c}_i$ are sampled sequentially from Multinomial distributions with probabilities $P_{ig}$. The addition, deletion, and update of the cluster allocation of the latent coordinates follow the following procedure. If, for the $i^{\text{th}}$ latent coordinate, we have that $\mathbf{c}_i = \bigl( c_{i1}, \dots, c_{ig}, \dots, c_{iG}, c_{i(G+1)} \bigr) =  \bigl(0, \dots, 0, \dots, 0, 1 \bigr) $, a new component is formed, which includes for now only $\mathbf{z}_i$. In this case, new component parameters need to be sampled from the corresponding full conditional distributions, previously defined. We propose to initialize first $\bm{\mu}_{G+1}$ with $\mathbf{z}_i$, then sample a value for $\bm{\Sigma}_{(G+1)}$ and update the value of $\bm{\mu}_{(G+1)}$. Also, to all the other $\bm{c}_j$ vectors, $j\neq i$ is added a $(G+1)^{\text{th}}$ $0$ element. If instead, $c_{ig} = 1$ for some $g\leq G+1$, the $(G+1)^{\text{th}}$ element is removed from vector $\bm{c}_i$, to uniform its length with those of other label vectors. Last, if the update of a vector $\bm{c}_i$ empties a given component $g$, then the $g^{\text{th}}$ element is removed from all $\bm{c}_i$ vectors, $i =1\dots,n$.

\subsection{Component-specific parameters}
\label{appendix:param}
The mixture component-specific parameters follow a Dirichlet Process, whose base distribution is a Normal Inverse Gamma (Section \ref{estimation}). For a given value of $G$, conjugacy allows a straightforward derivation of the full conditional distributions for $\bm{\mu}_g$ and $\bm{\Sigma}_g$, $g=1, \dots, G$:

\[
\mu_{rg} \mid \dots \thicksim {\cal N} \left( \frac{\tau_z \sum_{i \in g} z_{ri} + m_{rz}}{1 + n_g \tau_z},
 \frac{\tau_z \sigma_{rg}^2}{1+ n_g \tau_z} \right), \quad 
\sigma_{rg}^2 \mid \dots \thicksim  \mathrm{InvGamma} \bigl( x_g, X_{rg}\bigr), \quad r = 1, \dots p,
\]
with $n_g$ the number of latent coordinates currently assigned to the $g^{\text{th}}$ component, that is $n_g = \pi_g n$. Also,
\[
x_g = \frac{n_g +1+2\nu_1}{2}, \quad
X_{rg} = \frac{\tau_z \sum_{i \in g} (z_{ri} -\mu_{rg})^2+ (\mu_{rg} -m_{rg})^2 + 2\tau_z \nu_2}{2 \tau_z}.
\]
The components mean and covariance are then: 
\[
\bm{\mu}_{g} = \Bigl( \mu_{1g}, \dots, \mu_{pg} \Bigr)
\quad; \quad
\bm{\Sigma}_{g} = \begin{bmatrix}
    \sigma_{1g}^2     & \dots & 0 \\
    \vdots & \ddots & \vdots \\
    0    & \dots & \sigma_{pg}^2    \\
\end{bmatrix} \mathbf{I}, \quad
g = 1, \dots,G.
\]
If the hyperparameter $m_r$ is given standard Gaussian distribution, it may be updated from the following conditional distribution:
\[
m_{r}\mid \dots \thicksim {\cal N} \Biggl( \sum_{g=1}^G \frac{\mu_{rg}}{\sigma_{rg}^2}/ \bigl(\tau_z +x_r \bigr), \frac{\tau_z}{\tau_z + x_r} \Biggr), \quad r = 1, \dots, p,
\]
with $x_r= \sum_{g=1}^G \frac{1}{ \sigma_{rg}^2}$.

\subsection{Concentration parameter}
\label{appendix:concentration}
Following \cite{Escobar:1995}, the update of the Dirichlet process concentration parameter $\psi$ is performed as follows. Given a fixed value of $G$:
\begin{enumerate}
\item Simulate the latent variable $x \thicksim \mathrm{Beta}(\xi_1+1, n)$;
\item Compute $ \eta = \frac{s}{1+s}$, with $ s= \frac{\xi_1 +G -1}{n(\xi_2 -\log(x)}
$;
\item Generate a new value of the concentration parameter $\psi$ from: 
\[
\begin{cases}
\mathrm{Gamma} \Bigl( \xi_1 +G , \xi_2 -\log(x) \Bigr) \quad& \text{with probability} \quad \eta\\
\mathrm{Gamma} \Bigl( \xi_1 +G -1, \xi_2 -\log(x) \Bigr) \quad& \text{with probability} \quad 1-\eta
\end{cases}.
\]
\end{enumerate}

\section{Details of simulated data experiments}
\label{appendix:sim}
In all four simulated scenarios, we generate the logit parameters $\alpha^{(k)}$ and $\beta^{(k)}$ for all multiplex views from the following distributions:
\[
\alpha^{(k)} \thicksim \mathcal{N}_{[-0.5,\infty)}(1.5, (0.6)^2); \quad 
\beta^{(k)} \thicksim \mathcal{N}_{[0,\infty)}(0.7, (0.3)^2).
\]
In all four scenarios, the cluster means are simulated from a bivariate Gaussian distribution:
$$
\mu_g \thicksim \mathcal{N}_2(\mathbf{0}, (\log(G) + 1.5G)\mathbf{I}),
$$
for $G = \{2, 3, 4\}$. In Scenario I, II, IV, the cluster variances are generated from:
$$
\sigma_{rg}^2 \thicksim U((0.05)^2, (0.10)^2), \qquad r = 1, 2.
$$
In Scenario III, the cluster variances are generated from:
$$
\sigma_{rg}^2 \thicksim U((0.1)^2, (0.3)^2), \qquad r = 1, 2.
$$
\end{document}